\RequirePackage{amsmath}
\documentclass[runningheads,10pt]{llncs}
\usepackage[T1]{fontenc}
\usepackage{graphicx}
\usepackage{hyperref}
\usepackage[bb=boondox]{mathalfa}
\usepackage{multicol,multirow}
\usepackage{amsfonts}
\usepackage{amsmath}
\usepackage{algorithm}
\usepackage{xcolor, color, soul}
\sethlcolor{yellow}

\usepackage{algpseudocode}
\usepackage{tikz}
\usetikzlibrary{quantikz}
\usepackage{algpseudocode}
\usepackage{float}
\usetikzlibrary{matrix, shapes, arrows, positioning, chains, calc}
\usetikzlibrary{fit}
\usepackage{ulem}
\usepackage{adjustbox}

\usepackage{enumitem}
\usepackage{bigdelim}
\usepackage{subcaption}
\usepackage{bm}
\usepackage{verbatim}
\usepackage{orcidlink}

\usetikzlibrary{shapes, arrows, positioning}
\newlength{\nodedistance}

\usepackage[paragraphs, wordspacing, tracking, floats, mathspacing, subtle]{savetrees} 

\begin{document}
\newif\ifsubmission
\submissionfalse
\title{A practical key-recovery attack on LWE-based key-encapsulation mechanism schemes using Rowhammer}
\titlerunning{Rowhammer on Lattice-base KEMs}
%

%
%
\ifsubmission
\else
\author{Puja Mondal\inst{1}\orcidlink{0009-0006-7300-8435} \and
Suparna Kundu\inst{2}\orcidlink{0000-0003-4354-852X} \and
Sarani Bhattacharya\inst{3}\orcidlink{0000-0002-4190-2671} \and Angshuman Karmakar\inst{1,2}\orcidlink{0000-0003-2594-588X} \and Ingrid Verbauwhede\inst{2}\orcidlink{0000-0002-0879-076X}}
\authorrunning{P. Mondal et al.}
%
\institute{Department of Computer Science and Engineering, IIT Kanpur, India \\ \email{\{pujamondal,angshuman\}@cse.iitk.ac.in}\and
COSIC, KU Leuven, Kasteelpark Arenberg 10, Bus 2452, B-3001 Leuven-Heverlee, Belgium\\
\email{\{suparna.kundu,ingrid.verbauwhede\}@esat.kuleuven.be}\and
Department of Computer Science and Engineering, IIT Kharagpur, India\\
\email{sarani@cse.iitkgp.ac.in}
}

\fi

\maketitle              
\begin{abstract}
Physical attacks are serious threats to cryptosystems deployed in the real world. In this work, we propose a microarchitectural end-to-end attack methodology on generic lattice-based post-quantum key encapsulation mechanisms to recover the long-term secret key. Our attack targets a critical component of a Fujisaki-Okamoto transform that is used in the construction of almost all lattice-based key encapsulation mechanisms. We demonstrate our attack model on practical schemes such as Kyber and Saber by using Rowhammer. We show that our attack is highly practical and imposes little preconditions on the attacker to succeed. As an additional contribution, we propose an improved version of the plaintext checking oracle, which is used by almost all physical attack strategies on lattice-based key-encapsulation mechanisms. Our improvement reduces the number of queries to the plaintext checking oracle by as much as $39\%$ for Saber and approximately $23\%$ for Kyber768. This can be of independent interest and can also be used to reduce the complexity of other attacks.
\end{abstract}

\section{Introduction}
Post-quantum cryptography (PQC) refers to cryptographic protocols and algorithms designed to be secure against attacks by both classical and quantum computers. 
A large quantum computer can \textit{easily} subvert the security assurance of our current widely used public-key cryptographic (PKC) schemes based on integer factorization~\cite{RSA} and elliptic curve cryptography~\cite{ECC_miller_Crypto86} using Shor's~\cite{Shor_1994} and Proos-Zalka's~\cite{Proos_Zalka_2003} algorithm respectively. Therefore, it is imperative that we replace our existing PKC cryptographic with PQC schemes. However, the transition to post-quantum cryptography is a complex process that involves careful evaluation, standardization, and implementation of new cryptographic algorithms. A watershed moment in this process is the recently concluded standardization procedure by the National Institute of Standards and Technology (NIST)~\cite{nist_final_report}. NIST proposed the key-encapsulation mechanism (KEM) Kyber~\cite{KYBER} and digital signature schemes Dilithium~\cite{DILITHIUM}, Falcon~\cite{web:falcon} and SPHINCS+~\cite{web:sphincs} as PQC standards. 

Nevertheless, a pivotal step before a cryptosystem can be deployed for widespread public use is the assessment of its physical security. It is not a rare instance when the security of a mathematically secure cryptosystem is completely compromised by physical attacks~\cite{aranha_power_attack,DFA_EC,FA_RSA}. During the standardization process, NIST also highlighted resilience against physical attacks as one of the important criteria in the selection of standards. For physical security assessments, usually, two primary types of attacks are considered. First, passive side-channel attacks (SCA), that work by exploiting flaws in the implementation and using leakage of secret information through physical channels such as power consumption, electromagnetic radiation, acoustic channels, etc. Second, active fault attacks (FA), that work by disrupting the normal execution of a cryptographic scheme through laser radiation, power glitches, etc., and then manipulate the result of the faulty execution to extract the secret key. There exists another type of physical attack known as microarchitectural (MA) attacks. This type of attack exploits the vulnerabilities or imperfections in the architecture of the platform where the cryptographic scheme executes. A strong motivation for studying MA attacks is that while traditional side-channel and fault attacks primarily target small, low-power devices such as microcontrollers e.g. Cortex-M devices and Internet of Things (IoT) devices, MA attacks can affect a much broader range of platforms such as enterprise servers, cloud platforms where multiple honest processes share the same hardware with a potentially hostile process. The two former physical attacks require the attacker to have physical access to the target device, but MA attacks can be performed remotely. Also, there are some relatively simpler methods like constant-time coding techniques that can help defend against some side-channel attack vectors like simple-power analysis, but for MA attacks e.g. Rowhammer-induced bit-flips~\cite{Flippingbit1} cannot be easily mitigated through coding practices alone. In the past, successful MA attacks on classical cryptographic schemes such as elliptic-curve discrete signature algorithms and symmetric schemes such as AES~\cite{AES} have been demonstrated before~\cite{ladderleak,CacheAttack,CacheAttack1,CacheAttack2}.

Currently, there exist studies on physical attacks on PQC using SCA and FA~\cite{PesslP21,Sidechannelattacks1,DBLP:journals/iacr/MujdeiBBKWV22,HermelinkPP21} and some generic countermeasures such as masking and shuffling~\cite{Masked_kyber,proof_shuffle,HigherMasking}. At this moment there exists only a handful of MA attacks on PQ schemes such as digital signature schemes Dilithium and LUOV~\cite{DilithiumCore,LUOV} and key-encapsulation mechanism (KEM) Frodo~\cite{FRODOFLIP}. Among these only Dilithium is a PQC standard. Therefore, we can safely admit that currently there is a huge gap in the literature regarding the assessment and countermeasures of MA attacks.  As PQC seems to be prevalent in the near future it is crucial to study MA attacks in the context of PQC schemes. Hence, in this work, we focus on mounting efficient MA attacks on PQC schemes. We briefly summarize our contributions below.
\begin{itemize}[nosep]
    \item {We study MA attacks or specifically Rowhammer attacks on KEMs based on hard lattice problem learning with errors (LWE). Most LWE-based KEMs share a generic framework Lyubashevsky et al.~\cite{LPR} to first create public-key encryption and then convert it to key-encapsulation mechanism using a version of Fujisaki-Okamoto transform~\cite{FOT}. We sketch an outline of how such generic constructions can be attacked using a Rowhammer-based MA attack. In Rowhammer attacks an attacker repeatedly accesses the memory rows adjacent to the victim process's memory row. Such repeated access can result in bit-flips in the victim process's memory row. Rowhammer can be single-sided when the attacker accesses memory rows only on one side of the target memory row or more aggressively double-sided, where the attacker accesses memory rows above and below the target memory row. This happens due to imperfections in the dynamic random-access memory (DRAM). For interested readers, we have provided more details of Rowhammer in Appendix~\ref{app:rowhammer}.}
    \item {Physical attacks on chosen-ciphertext attack (CCA) secure KEMs~\cite{HermelinkPP21,PesslP21,DBLP:journals/iacr/MujdeiBBKWV22,Sidechannelattacks1} work by running the decapsulation procedure or the plaintext checking oracle multiple times with different ciphertexts. At each run side-channel traces are captured or faults are induced which reveal some part of the key. Therefore, reducing the number of invoking the plaintext checking oracle can make the attack more practical. The work in~\cite{ParallelPC1} proposed a method to reduce the number of times the plaintext checking oracle is invoked. Here, we further reduced the number of times the oracle is invoked by as much as $39\%$ for Saber and approximately $23\%$ for Kyber768 compared to the previous work using some offline computations. The advantage of our method is not limited to this work only and can be of independent interest in the context of physical attacks on lattice-based KEMs.}
    
    \item {We choose two PQ KEMs Kyber~\cite{KYBER} and Saber~\cite{SABER} to demonstrate the practicality of our attack. Kyber is a PQ KEM standard proposed by NIST and Saber was a finalist of the NIST PQC standardization procedure. We tailor our attack according to the design choices and parameters of Saber and Kyber.}
    \item {We show an end-to-end key-recovery method on Saber and Kyber based on remote software-induced faults only without using electromagnetic radiation, voltage glitch, laser radiation, etc. Our attack is very realistic as our conditions of attack are very relaxed compared to the previous works.  }
    \item {Finally, we discuss the effect of existing physical attack countermeasures on our attack.}
\end{itemize}

\ifsubmission

\else
\subsection{Paper Organization}

The structure of this paper is organized as follows. The paper is organized as follows:
Section 2 provides an overview of the necessary background information and introduces the notation and definitions used throughout the paper.
Section 3 reviews the previous research conducted in the field.
Section 4 presents the generic fault model of LPR schemes and explains its application to Kyber and Saber.
Section 5 focuses on the practical realization of the fault model.
\fi

\section{Preliminaries}

\noindent
\textbf{Notations: }We denote $\mathbb{Z}_{q}$ to represent the ring of integers modulo $q$. We use lowercase letters, lowercase letters with a bar, and uppercase letters to denote an element in $\mathbb{Z}_{q}$, vectors containing elements in $\mathbb{Z}_{q}$, and matrices with elements in $\mathbb{Z}_{q}$ respectively. Let $x\in \mathbb{Z}_{q},$ then $x^{i}$ represents the $i$-{th} bit of $x$. 
Bold lowercase letters are used to denote elements in $R_{q}$ where $R_{q}$ is the polynomial ring $\mathbb{Z}_{q}[x]/(x^{n}+1)$. For $i \in \{0,\ 1,\ \ldots,\ n-1\}$, ${\bf x}[i]$ represents the $i$-{th} coefficient of the polynomial ${\bf x}\in R_{q}$. $R_{q}^{l}$ represents the ring with vectors of $l$ polynomials of $R_q$ and the ring with matrices of $l\times k$ polynomials of $R_{q}$ is presented by $R_{q}^{l\times k}$. 
We use bold lowercase with a bar and bold uppercase letters to denote elements in $R_{q}^{l}$ and $R_{q}^{l\times k}$, respectively.
For ${\bf \Bar{x}}\in R_{q}^{l}$ and ${\bf X}\in R_{q}^{l\times k}$, ${\bf \Bar{x}}_{i}$ denotes the $i$-{th} polynomial of the vector ${\bf \Bar{x}}$ and ${\bf X}_{i,j}$ denotes the $(i,\ j)$-{th} polynomial of the matrix ${\bf X}$. The product of two polynomials ${\bf x}$ and $ {\bf y}$ is denoted by ${\bf xy}$. 
The inner product of ${\bf \Bar{x}}$ and ${\bf \Bar{y}}$ in $R_{q}^l$ is equal to $\sum_{i=0}^{l-1} {\bf \Bar{x}}_i {\bf \Bar{y}}_i$ is denoted by $\langle{\bf \Bar{x}},\ {\bf \Bar{y}}\rangle$.
If $x$ is sampled from the set $S$ according to the distribution $\mathcal{X}$, then we denote it as $x\leftarrow \mathcal{X}(S)$. We use $\mathcal{U}$ to represent uniform distribution and $\beta_{\nu}$ to indicate centered binomial distribution (CBD) with the standard deviation $\sqrt{\nu}/2$. $\lfloor x \rfloor$ outputs the largest integer, which is less than or equal to $x$. $\lfloor x\rceil$ represents the rounding of $x$ to the nearest integer, which is equal to $\lfloor x+\frac{1}{2}\rfloor$. $r \gg x$ and $r \ll x$ denotes $r$ shifted by $x$ bit positions towards right and left respectively. All these operations can be extended to the polynomials, vectors, and matrices by applying them coefficient-wise. The cardinality of a set $S$ is denoted by $|S|$.
\subsection{Learning with errors (LWE) problem and its variants}
\noindent \textbf{LWE problem:} Let us assume $ A\leftarrow \mathcal{U}(\mathbb{Z}_{q}^{l\times k})$, error $ \Bar{e}\leftarrow \chi(\mathbb{Z}_{q}^{l})$, secret $ \Bar{s}\leftarrow \chi(\mathbb{Z}_{q}^{k})$, $ \Bar{b} = A\Bar{s}+\Bar{e} \in \mathbb{Z}_{q}^{l}$, and $ \Bar{b}'\leftarrow \mathcal{U}(\mathbb{Z}_{q}^{l})$, where $l,\ k,\ n$ are positive integers and $\chi$ is a distribution. Then, the decision version of the LWE problem states that distinguishing between ${(A,\  \Bar{b})}$ and ${(A,\ \Bar{b}')}$ is hard. This hardness depends on the parameter $(n,\ l,\ k,\ q,\ \chi)$~\cite{OdedLecture}.  

\noindent \textbf{Ring-LWE (RLWE) problem:} If we use the polynomial ring $R_{q}=\mathbb{Z}_{q}[X]\big/(x^{n}+1)$ instead of $\mathbb{Z}_{q}$ and $l=k=1$, then we call the problem as Ring learning with error problem (RLWE)~\cite{LPR}. So, in the RLWE problem, given ${\bf a}\leftarrow \mathcal{U}(R_q)$, ${\bf e},\ {\bf s}\leftarrow \chi(R_q)$, ${\bf b = as+e} \in R_q$, and ${\bf b}'\leftarrow \mathcal{U}(R_q)$, it is hard to distinguish between ${(\bf a,\ \bf b)}$ and ${(\bf a,\ \bf b')}$. This hardness depends on the parameter $(n,\ q,\ \chi)$. 

\noindent \textbf{Module-LWE (MLWE) problem:} In the MLWE problem~\cite{MLWE}, ${\bf A}\leftarrow \mathcal{U}(R_q^{l\times l})$ and ${\bf \Bar{e}},\ {\bf \Bar{s}}\leftarrow \chi(R_q^{l})$, ${\bf \Bar{b} = A\Bar{s}+\Bar{e}} \in R_q^l$, and $\Bar{{\bf b}'}\leftarrow \mathcal{U}(R_q^l)$. The MLWE problem states that it is hard to distinguish between ${(\bf A,\ \Bar{b})}$ and ${(\bf A,\ \Bar{b'})}$. Here, the hardness depends on the parameter $(n,\ l,\ q,\ \chi)$. 

\noindent \textbf{Learning with Rounding (LWR) problem:} In this problem, the error sampling is replaced by the rounding operation. Let us assume ${\bf A}\leftarrow \mathcal{U}(\mathbb{Z}_{q}^{l\times k})$, ${\bf s}\leftarrow \chi(\mathbb{Z}_{q}^{k})$, $ {\bf b = \lfloor \frac{p}{q}(As) \rceil} \in \mathbb{Z}_{p}^{l}$, and ${\bf b}'\leftarrow \mathcal{U}(\mathbb{Z}_{p}^{l})$, where $q>p>0$. Then the LWR problem states that distinguishing between ${(\bf A,\ b)}$ and ${(\bf A,\ b')}$ is hard. This hardness depends on the parameter $(n,\ l,\ k,\ q,\ \chi)$~\cite{BanerjeePR12}.

The ring-LWR (RLWR) problem and the module-LWR (MLWR) problem can be defined from the LWR problem in a similar way as the RLWE problem and the MLWE problem are defined from the LWE problem.
\subsection{LPR public-key encryption}

\begin{figure}[b]
\centering
\fbox{\begin{varwidth}{0.98\textwidth}
\begin{subfigure}[t]{0.46\textwidth}
    \raggedright 
    \begin{small}
    $\mathtt{LPR}{.}\mathtt{PKE}{.}\mathtt{KeyGen} ()$
    \begin{enumerate}[wide=0em, itemsep=0pt, parsep=0pt, font=\scriptsize\tt\color{gray}]
            \item $\pmb{a} \leftarrow \mathcal{U}(R_q) $ \\
            \item $\pmb{s},\ \pmb{e} \leftarrow \mathcal{\chi}(R_q)$ \\
            \item $\pmb{b} =  \pmb{a}\pmb{s} + \pmb{e}$ \\
            \item \textbf{return} $(pk = (\pmb{a},\ \pmb{b}),\ sk =  (\pmb{a},\ \pmb{s}))$
    \end{enumerate} 
    \end{small}
\end{subfigure}
\begin{subfigure}[t]{0.46\textwidth}
    \raggedright
    \begin{small}
    $\mathtt{LPR}{.}\mathtt{PKE}{.}\mathtt{Enc} (pk = (\pmb{a},\ \pmb{b}),\ $ $\text{message } m \in \{0,\ 1\}^{n})$
    \begin{enumerate}[wide=0em, itemsep=0pt, parsep=0pt, font=\scriptsize\tt\color{gray}]
        \item $\pmb{r},\ \pmb{e_1},\ \pmb{e_2} \leftarrow \mathcal{\chi}(R_q)$ \\
        \item $\pmb{u} =  \pmb{a}\pmb{r} + \pmb{e_1}$
        \item $\pmb{v} =  \pmb{b}\pmb{r} + \pmb{e_2} + \mathtt{Encode}(m)$
        \item \textbf{return} $c =(\pmb{u}, \pmb{v})$
    \end{enumerate} 
    \end{small}
\end{subfigure}
\begin{subfigure}[t]{\textwidth}
    \raggedright
    \begin{small}
    $\mathtt{LPR}{.}\mathtt{PKE}{.}\mathtt{Dec}(sk=(\pmb{a},\ \pmb{s}),\ c=(\pmb{u},\ \pmb{v}))$
    \begin{enumerate}[wide=0em, itemsep=0pt, parsep=0pt, font=\scriptsize\tt\color{gray}]
        \item $\pmb{m}' = \pmb{v} - \pmb{u} \pmb{s}$
        \item $m = \mathtt{Decode} (\pmb{m}')$  \\
        \item \textbf{return} $m$
    \end{enumerate} 
    \end{small}
\end{subfigure}
\end{varwidth}}
\caption{CPA secure $\mathtt{LPR}{.}\mathtt{PKE}$~\cite{LPR}}
\label{fig:lprpke}
\end{figure}
Lyubashevsky, Peikert, and Regev proposed the LPR public-key encryption (PKE) scheme based on the RLWE problem~\cite{LPR} as shown in Figure~\ref{fig:lprpke}. 
Throughout this paper, we call this scheme as \texttt{LPR.PKE}. Here all the polynomials are elements of $R_{q}$, where $q$ is a prime number and $n$ is a power of two. In \texttt{LPR.PKE.KeyGen}, the secret ${\bf s} \leftarrow \chi (R_{q})$ and the error ${\bf e}\leftarrow \chi (R_{q})$.
Here, $\mathcal{\chi}$ is the Gaussian distribution, which is replaced by CBD in Kyber and Saber. 
${\bf a} \leftarrow \mathcal{U}(\mathbb{Z}_{q})$ and ${\bf b} = {\bf as+e} \in R_{q}.$ 
This algorithm declares $pk=({\bf a,\ b})$ as public key and saves $sk=({\bf a,\ s})$ as private key. In the \texttt{LPR.PKE.Enc} algorithm, a part of the ciphertext ${\bf u}$ is computed similarly to the public key ${\bf b}$. The other part of the ciphertext $\bf{v}$ contains message $m$ and is computed as ${\bf v=br+e_{2}}+\mathtt{Encode}(m)$. Here the $\mathtt{Encode}$ function is defined as $\mathtt{Encode}(m) = m \cdot \lfloor\frac{q}{2}\rfloor$ \textit{i.e.} multiplication of each message coefficient $m[i]$ with $\frac{q}{2}$. Then this algorithm outputs $c=({\bf u,\ v})$ as the ciphertext of the message $m$. The \texttt{LPR.PKE.Dec} algorithm takes the ciphertext $c$, and the secret key ${\bf s}$ as input, and then computes ${\bf m'=v-us}$. Now, 
\begin{align*}
    {\bf m'}&={\bf v-us} =({\bf br + e_{2} }+ \mathtt{Encode}(m)-{\bf (ar + e_{1})s}\\
    &={\bf (as+e)r+e_{2}} + m\cdot \lfloor{q/2}\rfloor-{\bf (ar + e_{1})s} =m\cdot \lfloor {q/2}\rfloor + {\bf er+e_{2}-e_{1}s}
\end{align*}
Here, ${\bf d= er+e_{2}-e_{1}s}$ is known as decryption noise. The \texttt{LPR.PKE.Dec} algorithm uses the $\mathtt{Decode}$ function to remove the decryption noise from the message polynomial $\bf m'$ and recovers the message $m\in\{0,\ 1\}^n$. 

\noindent
\textbf{Fujisaki-Okamoto (FO) transformation: } The \texttt{LPR.PKE} scheme provides security against chosen-plaintext attacks (CPA) but does not offer protection against chosen-ciphertext attacks (CCA). FO transform is a generic transform to transform a CPA-secure PKE to CCA-secure KEM. Due to the presence of noise in the LPR-based scheme, a variant of FO transformation proposed by Jiang et al.~\cite{Jiang2017} is generally used. 
The algorithms of this KEM are shown in Figure~\ref{fig:fo-kem}. A more detailed discussion regarding this FO transformation is provided in Appendix~\ref{app:fo}. 
\begin{figure}[b]
\centering
\fbox{\begin{varwidth}{0.98\textwidth}
\begin{subfigure}[t]{0.47\textwidth}
    \begin{small}
    \raggedright 
    $\mathtt{KEM}{.}\mathtt{KeyGen} ()$
    \begin{enumerate}[wide=0em, itemsep=0pt, parsep=0pt, font=\scriptsize\tt\color{gray}]
        \item $(pk,\ sk)  = \mathtt{PKE}{.}\mathtt{KeyGen} ()$ \\
        \item $z \leftarrow \mathcal{U}(\{0,\ 1\}^{n}$) \\
        \item \textbf{return} $(pk,\ {sk'} =  (sk || pk || \mathcal{H}(pk) || z)$
    \end{enumerate} 
    \end{small}
\end{subfigure}
\begin{subfigure}[t]{0.47\textwidth}
    \begin{small}
    \raggedright
    $\mathtt{KEM}{.}\mathtt{Encaps} (pk)$
    \begin{enumerate}[wide=0em, itemsep=0pt, parsep=0pt, font=\scriptsize\tt\color{gray}]
        \item $m  \leftarrow  \mathcal{U}(\{0,\ 1\}^{n}) $ \\
        \item $({K'},\ r) = \mathcal{G}(m,\ \mathcal{H}(pk))$ \\
        \item $ct = \mathtt{PKE}{.}\mathtt{Enc} (pk,\ m,\ r)$ \\
        \item $K = \mathcal{F}({K'},\ \mathcal{H}(ct))$
        \item \textbf{return} $(ct,\ K)$
    \end{enumerate} 
    \end{small}
\end{subfigure}
\begin{subfigure}[t]{\textwidth}
    \begin{small}
    \raggedright
    $\mathtt{KEM}{.}\mathtt{Decaps} ({sk'} = (sk || pk || \mathcal{H}(pk) || z),\ ct)$
    \begin{enumerate}[wide=0em, itemsep=0pt, parsep=0pt, font=\scriptsize\tt\color{gray}]
        \item $m'  =  \mathtt{PKE}{.}\mathtt{Dec} (sk,\ ct )$ \\
        \item $(K'',\ r') = \mathcal{G}(m',\ \mathcal{H}(pk))$ \\
        \item $c = \mathtt{LPR}{.}\mathtt{PKE}{.}\mathtt{Enc} (pk,\ m',\ r')$ \\
        \item \textbf{if: } $ct=c$ \textbf{return} $ K = \mathcal{F}({K''},\ \mathcal{H}(ct))$
        \item \textbf{else: } \textbf{return} $ K = \mathcal{F}(z,\ \mathcal{H}(ct))$  
    \end{enumerate} 
    \end{small}
\end{subfigure}
\end{varwidth}}

\caption{CCA secure $\mathtt{KEM}$ based on $\mathtt{LPR}{.}\mathtt{PKE}$ using FO transformation~\cite{Jiang2017}}
\label{fig:fo-kem}
\end{figure} 

\subsection{Kyber}
Kyber~\cite{KYBER} is an LPR-based KEM with MLWE as its underlying hard problem. In the key generation algorithm of Kyber, the secret ${\bf \Bar s\leftarrow \beta_{\eta_{1}}(R_{q}^{l})}$ and error ${\bf \Bar e}\leftarrow \beta_{\eta_{1}}(R_{q}^{l})$. One part of the public key ${\bf A}\leftarrow \mathcal{U}(R_{q}^{l\times l})$ and the another part of the public key is ${\bf \Bar b} = {\bf A}{\bf \Bar s}+{\bf \Bar e} $. The secret key is $sk=({\bf A},\ {\bf \Bar s})$. In the encryption algorithm, the errors ${\bf \Bar r}\leftarrow \beta_{\eta_{1}}(R_{q}^{l})$ and the errors ${\bf \Bar e}_{1}\leftarrow \beta_{\eta_{2}}(R_{q}^{l})$ and ${\bf e_{2}}\leftarrow \beta_{\eta_{2}}(R_{q})$. A part of the ciphertext ${\bf \Bar u}$ is computed similarly to the public key ${\bf \Bar b}$ generation. Another part of the ciphertext ${\bf v}= \langle{\bf \Bar b},\ {\bf \Bar r}\rangle +{\bf e_{2}}+\mathtt{Encode}(m)$, where $\mathtt{Encode}(m) = \lfloor m \cdot \frac{q}{2}\rfloor$. Then this algorithm uses $\mathtt{compress_q}$ to compress each coefficient of ${\bf \Bar u}$ to $d_{u}$ bits and ${\bf v}$ to $d_{v}$ bits. $c = ({\bf \Bar c}_{1},\ {\bf c_{2}}) = (\mathtt{compress_q}({\bf \Bar u}),\ \mathtt{compress_q}({\bf v},\ d_{v}))$ serves as the ciphertext associated with the message $m$. The decryption algorithm first decompresses both components ${\bf \Bar c}_{1}$ and ${\bf c_{2}}$ of the ciphertext $c$ with $\mathtt{Decompress_q}$ function. Suppose ${\bf \Bar u}'=\mathtt{Decompress_{q}}({\bf \Bar c}_{1},\ d_{u})$ and ${\bf v'}=\mathtt{Decompress_{q}}({\bf c_{2}},\ d_{v})$. Then it computes $\mathtt{Decode}({\bf v'} - \langle {\bf \Bar s}, \ {\bf \Bar u}'\rangle) = \mathtt{Compress_{q}}(({\bf v'} - \langle {\bf \Bar s},\  {\bf \Bar u}'\rangle),\ 1)$ to recover the message $m$. There are three security versions of Kyber based on the parameter set, and we include them in Table \ref{tab:parameters}. In this paper, unless otherwise specified we refer to the parameter set of Kyber$768$ with Kyber. For more details, we refer the interested reader kindly to the original paper~\cite{KYBER} for further details.

\begin{table}[t]
\centering
\caption{Parameter set of Kyber and Saber corresponding to different security levels}
\label{tab:parameters}
\resizebox{\textwidth}{!}{%
\begin{tabular}{|cl|ccccccc|c|c|c|}
\hline
\multicolumn{2}{|c|}{\multirow{3}{*}{\begin{tabular}[c]{@{}c@{}}Scheme\\ Name\end{tabular}}} & \multicolumn{7}{c|}{Parameters}                                                                                                                                                                                                                                                                                                       & \multirow{3}{*}{\begin{tabular}[c]{@{}c@{}}Post-quantum\\ Security\end{tabular}} & \multirow{3}{*}{\begin{tabular}[c]{@{}c@{}}Failure \\ Probability\end{tabular}} & \multirow{3}{*}{\begin{tabular}[c]{@{}c@{}}NIST\\ Security \\ Level\end{tabular}} \\ \cline{3-9}
\multicolumn{2}{|c|}{}                                                                       & \multicolumn{1}{c|}{\multirow{2}{*}{$l$}} & \multicolumn{1}{c|}{\multirow{2}{*}{$n$}} & \multicolumn{1}{c|}{\multirow{2}{*}{$q$}}      & \multicolumn{1}{c|}{\multirow{2}{*}{$p$}}      & \multicolumn{1}{c|}{\multirow{2}{*}{$T$}}  & \multicolumn{2}{c|}{\multirow{2}{*}{\begin{tabular}[c]{@{}c@{}}CBD\\ parameters\end{tabular}}} &                                                                                  &                                                                                 &                                                                                   \\
\multicolumn{2}{|c|}{}                                                                       & \multicolumn{1}{c|}{}                     & \multicolumn{1}{c|}{}                     & \multicolumn{1}{c|}{}                          & \multicolumn{1}{c|}{}                          & \multicolumn{1}{c|}{}                      & \multicolumn{2}{c|}{}                                                                          &                                                                                  &                                                                                 &                                                                                   \\ \hline
\multicolumn{1}{|c|}{\multirow{4}{*}{Kyber}}                   &                             & \multicolumn{1}{c|}{}                     & \multicolumn{1}{c|}{}                     & \multicolumn{1}{c|}{}                          & \multicolumn{1}{c|}{$p = 2^{d_u} $}            & \multicolumn{1}{c|}{$T = 2^{d_v} $}        & \multicolumn{1}{c|}{$\eta_1$}                            & $\eta_2$                            &                                                                                  &                                                                                 &                                                                                   \\ \cline{6-9}
\multicolumn{1}{|c|}{}                                         & Kyber512                    & \multicolumn{1}{c|}{2}                    & \multicolumn{1}{c|}{\multirow{3}{*}{256}} & \multicolumn{1}{c|}{\multirow{3}{*}{3329}}     & \multicolumn{1}{c|}{$2^{10}$}                  & \multicolumn{1}{c|}{$2^{4}$}               & \multicolumn{1}{c|}{3}                                   & 2                                   & $2^{107}$                                                                        & $2^{-139}$                                                                      & 1                                                                                 \\
\multicolumn{1}{|c|}{}                                         & Kyber768                    & \multicolumn{1}{c|}{3}                    & \multicolumn{1}{c|}{}                     & \multicolumn{1}{c|}{}                          & \multicolumn{1}{c|}{$2^{10}$}                  & \multicolumn{1}{c|}{$2^{4}$}               & \multicolumn{1}{c|}{2}                                   & 2                                   & $2^{166}$                                                                        & $2^{-164}$                                                                      & 3                                                                                 \\
\multicolumn{1}{|c|}{}                                         & Kyber1024                   & \multicolumn{1}{c|}{4}                    & \multicolumn{1}{c|}{}                     & \multicolumn{1}{c|}{}                          & \multicolumn{1}{c|}{$2^{11}$}                  & \multicolumn{1}{c|}{$2^{5}$}               & \multicolumn{1}{c|}{2}                                   & 2                                   & $2^{232}$                                                                        & $2^{-174}$                                                                      & 5                                                                                 \\ \hline
\multicolumn{1}{|c|}{\multirow{4}{*}{Saber}}                   &                             & \multicolumn{1}{c|}{}                     & \multicolumn{1}{c|}{}                     & \multicolumn{1}{c|}{$q = 2^{\epsilon_q} $}     & \multicolumn{1}{c|}{$p = 2^{\epsilon_p} $}     & \multicolumn{1}{c|}{$T = 2^{\epsilon_T} $} & \multicolumn{2}{c|}{$\mu$}                                                                     &                                                                                  &                                                                                 &                                                                                   \\ \cline{5-9}
\multicolumn{1}{|c|}{}                                         & LightSaber                  & \multicolumn{1}{c|}{2}                    & \multicolumn{1}{c|}{\multirow{3}{*}{256}} & \multicolumn{1}{c|}{\multirow{3}{*}{$2^{13}$}} & \multicolumn{1}{c|}{\multirow{3}{*}{$2^{10}$}} & \multicolumn{1}{c|}{$2^{3}$}               & \multicolumn{2}{c|}{5}                                                                         & $2^{107}$                                                                        & $2^{-120}$                                                                      & 1                                                                                 \\
\multicolumn{1}{|c|}{}                                         & Saber                       & \multicolumn{1}{c|}{3}                    & \multicolumn{1}{c|}{}                     & \multicolumn{1}{c|}{}                          & \multicolumn{1}{c|}{}                          & \multicolumn{1}{c|}{$2^{4}$}               & \multicolumn{2}{c|}{4}                                                                         & $2^{172}$                                                                        & $2^{-136}$                                                                      & 3                                                                                 \\
\multicolumn{1}{|c|}{}                                         & FireSaber                   & \multicolumn{1}{c|}{4}                    & \multicolumn{1}{c|}{}                     & \multicolumn{1}{c|}{}                          & \multicolumn{1}{c|}{}                          & \multicolumn{1}{c|}{$2^{6}$}               & \multicolumn{2}{c|}{3}                                                                         & $2^{236}$                                                                        & $2^{-165}$                                                                      & 5                                                                                 \\ \hline
\end{tabular}%
}
\end{table}
\subsection{Saber}
Saber~\cite{SABER} is a KEM based that also follows the LPR model. Saber is based on the hard problem MLWR. Here $q$ in $R_q$ is power-of-two. In the key generation algorithm of Saber, the secret ${\bf \Bar s} \leftarrow \beta_{\mu} (R_{q}^{l})$. The public key here is $({\bf A},\ {\bf \Bar b})$ where ${\bf A}\in R$ is an element of $R_{q}^{l\times l}$ and is sampled uniformly and ${\mathbf{ \Bar b}} = ({\mathbf{ A}}{\mathbf {\Bar s}}+{\mathbf{ \Bar h}})\gg({\epsilon_{q}-\epsilon_{p}})\in R_{p}^{l} $. The vector $\bf{ \Bar h}$ is needed for rounding, and it consists of constant polynomials with each coefficient equal to $2^{\epsilon_{q}-\epsilon_{p}-1}$. In the case of the encryption algorithm, $\bf{ \Bar s}'$ is also sampled from the CBD distribution $\beta_{\mu}$. The key contained part of the ciphertext $\bf{ \Bar u}$ is computed similarly to the public key $\bf{ \Bar b}$. The message contained part of the ciphertext ${\bf v}$ is computed as $(\langle {\bf \Bar b},\ {\bf \Bar s}'\rangle+{\bf h_{1}}-\mathtt{Encode}(m) \bmod p)\gg{\epsilon_{p}-\epsilon_{T}}\in R_{T}$. $\bf h_{1}$ is a constant polynomial with each coefficient equal to $2^{\epsilon_{q}-\epsilon_{p}-1}$. It is required for rounding. Let $c=(\bf{ \Bar u},\ {\bf v})$ is the ciphertext corresponding to the message $m$. Then, the decryption algorithm takes the ciphertext $c=(\bf{ \Bar u},\ {\bf v})$ and secret $\bf{ \Bar s}$ as inputs. It computes $({\langle{ \bf \Bar u},\ {\bf \Bar s}\rangle \bmod p}-2^{ \epsilon_{p}-\epsilon_{T}}{\bf v} + {\bf h_{2}} ) \bmod p\gg (\epsilon_{p}-1)\in R_{2}$ to find the decrypted message. $\bf h_{2}$ is also a constant polynomial with each coefficient equal to $2^{\epsilon_{p}-2}-2^{\epsilon_{p}-\epsilon_{T}-1}$. Like Kyber, Saber also has three security versions depending on the parameter set, and we present them in Table \ref{tab:parameters}. Similar to Kyber, in this paper, we refer to the parameter set of Saber with $l=3$ with Saber, and we refer to the original paper~\cite{SABER} for further details.

\subsection{Related works}

Lattice-based post-quantum KEMs are vulnerable to side-channel attacks. A timing attack on the \texttt{KEM.Decaps} has been shown in~\cite{TimingPQC}, it targets the non-constant time implementation of the ciphertext equality checking (Line 4 in \texttt{KEM.Decap} algorithm of Figure~\ref{fig:fo-kem}). \cite{Sidechannelattacks1}, proposed a generic and practical  Electromagnetic (EM) power analysis assisted CCA on LWE-based KEMs. They also target the \texttt{KEM.Decaps} in their attack. They have constructed a plaintext-checking oracle $\mathcal{O}$ with the help of an EM power attack, which can distinguish two particular messages $m_{1}=00\dots 0 \text{ (all zeros)}$ and $m_{2}=00\dots 01 \text{ (all zeros except the LSB)}$. This oracle provides single-bit information related to one coefficient of the secret key. Continuing the same methods, the attacker can find the whole secret key. This paper has shown that $2000$ to $4000$ queries are required to retrieve the complete secret key for Kyber. \cite{ParallelPC} reduced the query requirements by creating a multiple-valued plaintext-checking oracle. Here, the attacker acquires information regarding multiple secret key coefficients from a single query. In~\cite{ParallelPC1}, the authors further reduced the number of queries required to recover the whole secret keys by improving the model of plaintext-checking oracle. One significant area of research in this domain revolves around improving the efficiency of attacks by minimizing the number of required quires. This reduction enables a more precise evaluation of the cost of an optimal attack. Our attack contributes in this direction by improving the process of using the parallel plaintext checking oracle model of the paper~\cite{ParallelPC1}.

Rowhammer has been used to successfully attack many cryptographic primitives. In the paper \cite{RSASig}, researchers demonstrated a Rowhammer attack on RSA signatures. Additionally, in \cite{ReadBit}, the authors illustrate the direct reading of RSA key bits from the memory address. However, there is limited research on Rowhammer attacks targeting post-quantum schemes. The current state-of-the-art in this domain focuses on a single work involving Rowhammer attacks on the PQC KEM Frodo~\cite{frodo_scheme}. This research primarily targets the key-generation procedure, which is known to be relatively easy to protect. In this work, we will demonstrate an end-to-end Rowhammer attack on the decapsulation algorithm of the targeted schemes.


\section{Our attack using binary decision tree on the LPR-based schemes} \label{sec:attack_surface}

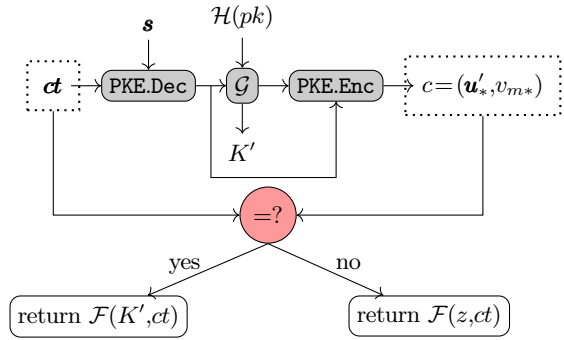
\begin{figure}[t]
     \centering
\setlength{\nodedistance}{4mm}
\begin{tikzpicture}[ 
node distance = \nodedistance,
operation/.style = {circle, draw=black},
bigoperation/.style = {rounded corners, draw=black},
operationsecret/.style = {circle, draw=black, fill=red!40!white, text=black},
bigoperationsecret/.style = {rounded corners, draw=black, fill=black!20!white, text=black},
operant/.style = {},
halfblockdraw/.style = {draw, rounded corners},
     line/.style = {draw, -Latex}
                       ]
\node (multbps) [bigoperationsecret]  {$\mathtt{PKE.Dec}$};
\node (bp) [operant, left=of multbps.west]  {$\pmb{ct}$};
\node (s) [operant, above=of multbps.north]  {$\pmb{s}$};
\draw[->] (bp) -- (multbps);
\draw[->] (s) -- (multbps);

\node (g) [bigoperationsecret, right=of multbps]  {$\mathcal{G}$};
\coordinate  (m) at ($(multbps.east)!0.5!(g.west)$);
\node (pkh) [operant, above=of g]  {$\mathcal{H}(pk)$};
\draw[->] (pkh) -- (g);
\draw[->] (multbps) -- (g);
\node (K) [operant, below=of g]  {$K'$};
\draw[->] (g) -- (K);

\node (beta1) [bigoperationsecret, right=of g]  {$\mathtt{PKE.Enc}$};
\draw[->] (g) -- (beta1);
\coordinate [below=0.25\nodedistance of K](tmp);
\draw[->] (m) -- (m |- tmp) -- (beta1 |- tmp) -- (beta1);
\node (beta2) [operant, right=of beta1]  {$c=(\pmb{u}'_*, {v}_{m*})$};
\draw[->] (beta1) -- (beta2);

\node[draw=black, dotted, thick, fit=(beta2)](ciphertext2) {};
\node[draw=black, dotted, thick, fit=(bp)](ciphertext) {};
\coordinate [below=1.5\nodedistance of K](x); 
\coordinate (y) at ($(ciphertext)!0.5!(ciphertext2)$);

\node (comp) [operationsecret] at (y |- x)  {$=?$};
\node (correct) [bigoperation, below left=2\nodedistance and 2\nodedistance of comp]  {return $\mathcal{F}(K', ct)$};
\draw[->] (comp.south) -- (correct);
\node[operant,  below left=0.5\nodedistance and 1.3\nodedistance of comp] (yes) {yes};

\node (correct) [bigoperation, below right=2\nodedistance and 2\nodedistance of comp]  {return $\mathcal{F}(z, ct)$};
\draw[->] (comp.south) -- (correct);
\node[operant,  below right=0.5\nodedistance and 1.3\nodedistance of comp] (no) {no};

\draw[->] (ciphertext) -- (ciphertext |- x) -- (comp);
\draw[->] (ciphertext2) -- (ciphertext2 |- x) -- (comp);
\end{tikzpicture}

\caption{Decapsulation algorithm of KEM based on \texttt{LPR.PKE} (Figure \ref{fig:fo-kem}). Here $z$ is a random number generated in the \texttt{KEM.KeyGen()} algorithm (Figure~\ref{fig:fo-kem}). The fault location is marked in red.}
\label{fig:LPRDecap}
\end{figure}

\noindent
\textbf{Attack Surface: }The KEM based on \texttt{LPR.PKE} shown in Figure \ref{fig:fo-kem} is resistant to CCA. In such schemes, the secret key is generated using the \texttt{KEM.KeyGen} is \textit{non-ephemeral} \textit{i.e.} stored and used for the long term. The key generation and encapsulation processes are executed only once. Therefore, the attacker needs to recover the secret key or the shared key from a single execution. However, the secret key remains fixed in the decapsulation algorithm for a long time and is used to derive the shared secret key $K$ from multiple users. This is done to remove the huge overhead of running the key generation process and distributing the public key each time two communicating parties want to establish the shared secret $K$. However, this convenience also helps an attacker.  An attacker can now execute the decapsulation operation multiple times and collect multiple traces or induce faults at different locations. This helps the attacker to refine its attack strategy and increase the probability of success manifold. This is why attacking the decapsulation operation is mostly chosen by attackers to mount physical attacks~\cite{PesslP21,HermelinkPP21,DBLP:journals/iacr/MujdeiBBKWV22,Sidechannelattacks1}. So, we also choose the decapsulation method as our target. The structure of the \texttt{KEM.Decaps} given in Figure~\ref{fig:fo-kem} is shown in Figure~\ref{fig:LPRDecap}. Here we also assume the attacker can invoke the victim's decapsulation procedure by submitting any ciphertexts of its preference. 

\ifsubmission
\else
We assume the general Rowhammer threat model, where the attacker and victim use two different processes in the same operating system or two virtual machines on the same server~\cite{XiaoZZT16}. This threat model is also used in most of the micro-architectural attacks work~\cite{YaromF14}. Here the attacker shares the same hardware responsible for performing the victim's decapsulation procedure of LPR-based KEM. The attacker can also invoke the victim's decapsulation procedure by submitting any ciphertexts of its preference. 
\fi



\subsection{Implementing a parallel plaintext checking (PC) oracle.}

In the \texttt{KEM.Decaps} procedure in Figure~\ref{fig:fo-kem}, the decrypted message $m$ undergoes a hashing operation $\mathcal{G}$ with the public key.
The resulting hash, denoted as $(K',\ r)$, where $r$ is combined with the message $m$ and is used as input for the subsequent re-encryption procedure using \texttt{LPR.PKE.Enc} algorithm. The generated key $K'$ is employed to create a valid shared key $K$. It is crucial to note that the hash function $\mathcal{G}$ is deterministic and solely relies on the decrypted message $m$ and public key $pk$. Considering $2^{t}$ messages where a fixed chunk of $t$ bits are changed while keeping all other $n-t$ bits fixed, such as 
\begin{align*}
\label{messages}
m^{(0)}&=\underbrace{000\dots0}_{t~bits}\underbrace{000\dots0}_{(n-t)~bits}\\ 
m^{(1)}&=\underbrace{100\dots0}_{t~bits}\underbrace{000\dots0}_{(n-t)~bits}\\
m^{(2)}&=\underbrace{010\dots0}_{t~bits}\underbrace{000\dots0}_{(n-t)~bits}\\
\dots\\
m^{(2^{t}-1)}&=\underbrace{111\dots1}_{t~bits}\underbrace{000\dots0}_{(n-t)~bits}
\end{align*}
A variation of $t$ bits in these messages leads to substantial variations in the computations performed during the hash $\mathcal{G}$ operation. Consequently, the ciphertexts generated by the \texttt{LPR.PKE.Enc} algorithm will differ for each of the $2^{t}$ messages.

In our attack scenario, we require the output to be dependent on the decrypted message. However, if we use artificially constructed ciphertext $ct$ (which is not generated from \texttt{LPR.PKE.Enc}), then with high probability, the re-encrypted ciphertext $c$ and $ct$ will be unequal. The current implementation always returns $\mathcal{F}(z,\ \mathcal{H}(ct))$ as the shared key, which is independent of the decrypted message. In order to distinguish the potential $2^{t}$ decrypted messages of the ciphertext $ct$, we need the output to be message-dependent. By omitting this equality checking condition, we ensure that the hash value $\mathcal{F}(K',\ \mathcal{H}(ct))$ is consistently returned, which is decrypted message dependent. That allows us to differentiate between the possible decrypted messages of $ct$. Our goal is to reliably acquire the shared key $\mathcal{F}(K',\ \mathcal{H}(ct))$ by employing a physical attack. 

In the \texttt{KEM.Decaps}, both Saber and Kyber use a variable named "fail". Compare the ciphertexts $ct$ and $c$ by calling the function \texttt{verify(c, ct, BYTES\_CCA\_DEC)} and storing the return value of this function in the "fail" variable. If the value of fail is $0$, then it returns the shared key $\mathcal{F}(K',\ \mathcal{H}(ct))$, which depends on the decrypted message. Otherwise, it returns the random shared key. Our aim is to 
flip the value of the variable "fail" by introducing fault even when the ciphertexts are not equal.

\subsection{Generic attack model using PC oracle}

The first stage of our attack is to carefully craft ciphertexts $c$ to reduce the number of invocations of the \texttt{KEM.Decaps} procedure. 
Here, we target to recover $t$ secret coefficients of the secret key $\mathbf{s}$ at a time. We introduce a notation $s_{i}^{(t)}$ to represent a block of consecutive $t$ coefficients of $\mathbf{s}$, where $i \in \{0,\ 1,\ \ldots, \lfloor\frac{n}{t}\rfloor-1\}$ and the last block $s_{\lfloor\frac{n}{t}\rfloor}^{(t')}$ consists $t'=n-(\lfloor\frac{n}{t}\rfloor\times t)$ secret coefficients. This ciphertext $c$ is then transmitted to the oracle $\mathcal{O}_{\mu}$. Here the oracle $\mathcal{O}_{\mu}$ defined as follows:
\begin{equation*}
{\mathcal{O}_{\mu}}( c;\ x^{(0)},\ x^{(1)},\ldots,\ x^{(\mu-1)})= r, \text{ if } {\tt PKE.Dec}(c)=x^{(r)}, 0\leq r\leq \mu-1
\end{equation*}. This oracle $\mathcal{O}_{\mu}$ takes a ciphertext $c$ and $\mu$ number of messages $x^{(i)}$ and returns the value $r$ such that the decrypted message of $c$ is $x^{(r)}$.
Upon receiving the ciphertext, the oracle $\mathcal{O}_{\mu}$ processes $c$ along with a set of potential messages $x^{(0)},\ x^{(1)},\ldots,\ x^{(\mu-1)}$. Then, the oracle provides a response $r$ such that $\mathtt{LPR.PKE.Dec}(sk,\ c)=x^{(r)}$. By analyzing the decrypted message $x^{(r)}$, we gain knowledge about the secret block $s_{i}^{(t)}$. As each secret coefficient is intricately tied to the decrypted message, this process gradually reduces the dimension of the secret coefficients within the targeted block. This reduction process involves considering the relationship between the decrypted message and the secret coefficients. After successfully reducing (not fully recovering) the dimension of the secret block, we construct another new ciphertext $c_{\alpha}$ that exploits the potential secret block $s_{i}^{(t)}$. Then, repeating the aforementioned process, we further reduce the cardinality of the secret set corresponding to each coefficient of the secret block to get our desired secret. The challenge lies in determining how many iterations of this process are necessary to effectively reduce the dimension of the secret block $s_{i}^{(t)}$. One possible approach is to repeat until the entire secret block $s_{i}^{(t)}$ is obtained. In the paper~\cite{ParallelPC1}, the authors used this approach. In this method, we need to query the oracle $\mathcal{O}_{\mu}$ $\lceil \log|S_{0}|\rceil$ times to find each of the secret blocks $s_{i}^{(t)}$ and $s_{\lfloor\frac{n}{t}\rfloor}^{(t')}$, where $i \in \{0,\ 1,\ \ldots,\ \lfloor\frac{n}{t}\rfloor-1\}$ and $t'=n-(\lfloor\frac{n}{t}\rfloor\times t)$. Here, $S_{0}$ represents the set of all possible values of a coefficient of the secret key. However, each iteration incurs a cost regarding the number of injected faults. Since each fault is resource-intensive, the objective is to find the secret with the minimum number of faults.

In our approach, we reduce the number of queries to the oracle $\mathcal{O}_{\mu}$ to find the all secret blocks $s_{i}^{(t)}$ and $s_{\lfloor\frac{n}{t}\rfloor}^{(t')}$, where $i \in \{0,\ 1,\ \ldots, \lfloor\frac{n}{t}\rfloor-1\}$ and $t'=n-(\lfloor\frac{n}{t}\rfloor\times t)$. Here, the previous approach is repeated $\lfloor \log|S_{0}|\rfloor$ times to progressively reduce the cardinality of the secret set corresponding to each coefficient of each secret block. Since we query $\lfloor \log|S_{0}|\rfloor$ times to the oracle $\mathcal{O}_{\mu}$ for each block, there will be some secrets that have not been determined yet. So, after reducing the dimension of each secret block, an index set, denoted as \texttt{Index[]}, is created to track the indices of the secret coefficients that have not been determined yet. A new ciphertext $c_{\alpha}$ is then constructed based on the \texttt{Index[]} set, and the values of the secret coefficients corresponding to the indices in \texttt{Index[]} are updated accordingly. For simplicity, we describe this attack template step by step for a parallelization factor $t$, which is a divisor of $n$, to unveil the secret block gradually. The process will be similar for other parallelization factor $t$.

\subsubsection{Constructing the ciphertext $c$}
Here, we present a method to construct a dummy ciphertext $ct = ({\bf u},\ {\bf v})\in R_{q}\times R_{q}$. This method helps to decrease the number of queries required to retrieve all the secrets of the block $s_{0}^{(t)}$, which contains $\{\mathbf{s}[0],\ \mathbf{s}[1],\ \ldots,\ \mathbf{s}[t-1]\}$ first $t$ coefficients of the secret polynomial $\mathbf{s}$. To construct the ciphertext $ct$, first we set $\mathbf{u}[0] = k_{u}$ and $ \mathbf{v}[j] = k_{v_{j}} $, $\forall 0\leq j\leq t-1$ are non zero and others coefficients of ${\bf u}$ and ${\bf v}$ are zero.  
Then \begin{equation*}
    {\bf (v-us)}=\sum_{j=0}^{t-1}k_{v_{j}}.x^{j}-\sum_{j=0}^{n-1} k_{u}\mathbf{s}[j].x^{j}
    \end{equation*}
\begin{equation*}
    \text{So } {\bf (v-us)}[j]=\begin{cases}
        (k_{v_{j}} - k_{u}\mathbf{s}[j]) & \text{if } 0\leq j\leq t-1\\
        (- k_{u}\mathbf{s}[j]) & \text{Otherwise }\,.
    \end{cases}
\end{equation*}

\noindent Hence the coefficients of the decrypted message $m$ will be 
\begin{equation*}
m^{j}=
\begin{cases}
    \mathtt{Decode}(k_{v_{j}} - k_{u}\mathbf{s}[j]) & \text{if } 0\leq j\leq t-1\\
    \mathtt{Decode}(-k_{u}\mathbf{s}[j]) & \text{Otherwise }\,. 
\end{cases}    
\end{equation*}
\noindent We choose the value $(k_{u},\ k_{v_{0}},\ k_{v_{1}},\ldots,\ k_{v_{t-1}} )$ such that 
\begin{equation*}
m^{j}=
    \begin{cases}
        \text{Depends on }\mathbf{s}[j] & \text{if } 0\leq j\leq t-1\\
        0 & \text{Otherwise } 
    \end{cases}    
\end{equation*}
We construct a binary decision tree shown in Figure~\ref{generic} to distinguish the secrets. We select each value $k_{v_{j}}$ from the tree accordingly.
Initially, all the values $k_{v_{j}}$ will be the root value $d_{0}$. Then, depending on the decrypted message, we update the value $k_{v_{j}}$ from the tree. Also, the value of $k_{u}$ will be fixed in an iteration because we are constructing the dummy ciphertext to get $t$ bits of information at a time.

To recover $j'$-{th} secret block, ${\bf s}_{j'}^{(t)}$ that contains the secret coefficients ${\bf s}[j'],\ {\bf s}[j'+1],\ \ldots,\ {\bf s}[j'+t-1]$), where $j'>0$ we have to construct the dummy ciphertext $ct = ({\bf u},\ {\bf v})\in R_{q}\times R_{q}$, where ${\bf u}[n-j'] = k_{u}$ and $ {\bf v}[j] = k_{v_{j}} $, $\forall 0\leq j\leq t-1$ are non zero and others coefficients of ${\bf u}$ and ${\bf v}$ are zero. Then 
\begin{equation*}
    {\bf (v-us)}=\sum_{j=0}^{t-1}k_{v_{j}}.x^{j}+\sum_{j=j'}^{n-1} k_{u}\mathbf{s}[j].x^{j-j'}-\sum_{j=0}^{j'-1} k_{u}\mathbf{s}[j].x^{n-j'+j}\\
\end{equation*}
\noindent Here, the decrypted message $m$ will be 
\begin{equation}
\label{gen_cipher}
m^{j}=
    \begin{cases}
        \mathtt{Decode}(k_{v_{j}} + k_{u}\mathbf{s}[j'+j]) & \text{if } 0\leq j\leq t-1\\
        \mathtt{Decode}(-k_{u}\mathbf{s}[j]) & \text{Otherwise } 
    \end{cases}    
\end{equation}

Similarly, the value of $k_{v_{j}}$ will be taken from the binary decision tree pictured in Figure~\ref{generic}. We also present the algorithm to create ciphertext in Algorithm~\ref{create_cipher}. 

\begin{algorithm}[b]
\caption{Ciphertext creation I}\label{create_cipher}
 \hspace*{\algorithmicindent} \textbf{Input:} The index $i$ of secret block $s_{i}^{(t)}$ and the current secret set $S_{r_{k}}$ corresponding to the block.\\
 \hspace*{\algorithmicindent}\textbf{Output:}  Ciphertext $ct$ such that the decrypted message of $c$ will be zero except the first $t$ positions.
	\begin{algorithmic}[1]
            \For{$k=0; k<n; k++$}
                \State ${\bf u}[k]=0;$ ${\bf v}[k]=0$
            \EndFor
            \State ${\bf u}[(n-i)\%n]=k_{u}$
            \For{$k=0; k<t; k++$}
                \If{${\bf s}[i+k]\in S_{r_{k}} $}
                \State ${\bf v}[k]=d_{r_{k}}$
                \EndIf
            \EndFor
\end{algorithmic}
\end{algorithm}
\subsubsection{Parallel PC oracle for $s_{i}^{(t)}$ by pruned binary decision tree:}
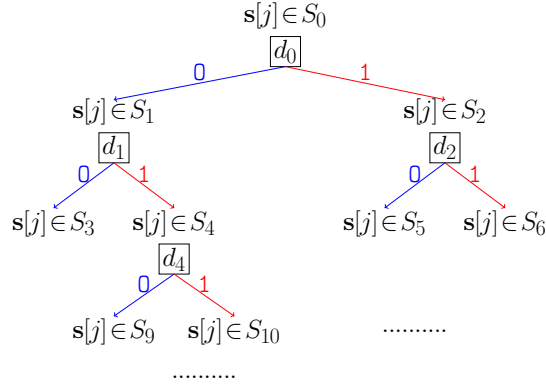
\begin{figure}[t]
	\centering
	\scalebox{.4}{
		\begin{tikzpicture}
			\node[font=\Huge](a) at (5.7,10){${\bf s}[j]\in S_{0}$};
			\node[ font=\Huge,rectangle, draw, minimum size=1cm] (a) at (5.7,8.8) {$d_{0}$};
			\node[font=\Huge](b) at (0,6.8){${\bf s}[j]\in S_{1}$};
			\node[font=\Huge](b) at (11,6.8){${\bf s}[j]\in S_{2}$};
			\draw[font=\Huge][blue,->](5.7,8.3) -- (0,7.2)node[midway,above]{{\tt O} };
			\draw[font=\Huge][red,->](5.7,8.3) -- (11,7.2)node[midway,above]{{\tt 1} };
			\node[font=\Huge ,rectangle, draw,minimum size=1cm] (a) at (11,5.6) {$d_{2}$};
			\node[font=\Huge](b) at (9,3.1){${\bf s}[j]\in S_{5}$};
			\node[font=\Huge](b) at (13,3.1){${\bf s}[j]\in S_{6}$};
			\draw[font=\Huge][blue,->](11,5.1) -- (9,3.6)node[midway,above]{{\tt O} };
			\draw[font=\Huge][red,->](11,5.1) -- (13,3.6)node[midway,above]{{\tt 1} };
			\node[font=\Huge, rectangle, draw,minimum size=1cm] (c) at (0,5.6) {$d_{1}$};
			\node[font=\Huge](b) at (-2,3.1){${\bf s}[j]\in S_{3}$};
			\node[font=\Huge](b) at (2,3.1){${\bf s}[j]\in S_{4}$};
			\draw[font=\Huge][blue,->](0,5.1) -- (-2,3.6)node[midway,above]{{\tt O} };
			\draw[font=\Huge][red,->](0,5.1) -- (2,3.6)node[midway,above]{{\tt 1} };
			\node[ font=\Huge ,rectangle, draw,minimum size=1cm] (a) at (2,1.9) {$d_{4}$};
			\node[ font=\Huge](c) at (0,-0.5){${\bf s}[j]\in S_{9}$};
			\node[ font=\Huge](c) at (4,-0.5){${\bf s}[j]\in S_{10}$};
			\draw[ font=\Huge][blue,->](2,1.4) -- (0,0)node[midway,above]{{\tt O} };
			\draw[ font=\Huge][red,->](2,1.4) -- (4,0)node[midway,above]{{\tt 1} };
            \node[ font=\Huge](c) at (10,-0.5){$..........$};
            \node[ font=\Huge](c) at (3,-2){$..........$};
		\end{tikzpicture}
	}
	\caption{Binary tree to select the value of $k_{v_{j}}=d_{i}$ for each $v[j]$}
	\label{generic}
\end{figure}
We construct a binary decision tree with two types of nodes; one is the $(S_{y},\ d_{y})$ where secret set $S_{y}$ with $|S_{y}|>1$ and the constant values $d_{y}$ which helps us to split the secret set $S_{y}$ into two disjoint sets $S_{2y+1}$ and $S_{2y+2}$. The other one is $S_{y}$ with $|S_{y}|=1$ as shown in Figure~\ref{generic}. We construct the tree such that the tree will be almost complete and the distance of node $(S_{y},\ d_{y})$ from the root node $(S_{0},\ d_{0})$ will be longer if the set $S_{y}$ contains the secret coefficients with comparatively lower probability. Let $h$ be the maximum height of the node of format $(S_{y},\ d_{y})$ from the root node $(S_{0},\ d_{0})$. Without loss of generality, assume that this maximum height node is $(S_{w},\ d_{w})$ i.e., the distance of the node $(S_{w},\ d_{w})$ from the root node is $h$ and the height of the tree is $h+1$ which is the distance from the node $(S_{0},\ d_{0})$ to the node $S_{2w+1}$. We have distinguished the secrets from the tree as follows:

First, we will query to the oracle $\mathcal{O}_{\mu}$ with the constructed ciphertext $ct$ and $2^{t}$ messages $m^{(0)},\ m^{(1)},\ldots,\ m^{(2^{t}-1)}$ described before. Let $m^{(r)}$ be the decrypted message of the ciphertext $ct$, which is received from the oracle $\mathcal{O}_{\mu}$. If the $j$-{th} secret coefficient of the block ${\bf s}_{i}^{(t)}$, ${\bf s}[i+j]\in S_{y}$ and $|S_{y}|>1$, then we will distinguish ${\bf s}[i+j]\in S_{2y+1}$ or ${\bf s}[i+j]\in S_{2y+2}$ according to the value of the corresponding $j$-{th} message bit of $m^{(r)}$ which is $\mathtt{Decode}(d_{y} - k_{u} {\bf s}[i+j])=0/1$ i.e., observing the current secret set $S_{y}$ in which the secret coefficient belongs and the decrypted bit $\mathtt{Decode}(d_{y} - k_{u} {\bf s}[i+j])$, we reduce the possible values of the secrets from $S_{y}$ to $S_{2y+1}$ or $S_{2y+2}$. In each iteration of the block ${\bf s}_{i}^{(t)}$, each value of $k_{v_{j}}$ will traverse this tree from the root node $(S_{0},\ d_{0})$ (with height $0$). In our attack, we traverse each value $k_{v_{j}}$ from the tree up to the height $h-1$, i.e., We pruned the highest heighted node $(S_{w},\ d_{w})$ from this tree. In this way, we reduce the cardinality of the secret set corresponding to each secret. Since we ignore the highest height node $(S_{w},\ d_{w})$, only secret coefficients that belong to the secret set $S_{w}$ will still be undetected.

\begin{algorithm}[b]
	\caption{Cardinality reduction of the secret set of the block ${s}_{i}^{(t)}$ }\label{decre_secret}
  \hspace*{\algorithmicindent} \textbf{Input:} The decrypted        
             message $m$ of the ciphertext $c$ such that $m$ is non-zero at most in the first $t$ positions.\\
    \hspace*{\algorithmicindent} \textbf{Input:} The value $r_{k}$ such that $\mathbf{s}[i+k]\in S_{r_{k}}$, $0\leq k\leq t-1$.\\
 \hspace*{\algorithmicindent}\textbf{Output:}  Update $[i+k]$ where $0\leq k\leq t-1$. 
	\begin{algorithmic}[1]
            \For{$l=0; l<t; l++$}
                \If{$m^{l}=0 $}
                \State $\mathbf{s}[i+l]\in S_{2r_{l}+1}$
                \Else 
                \State $\mathbf{s}[i+l]\in S_{2r_{l}+2}$
                \EndIf
            \EndFor
\end{algorithmic}
\end{algorithm}

\subsubsection{Construction of \texttt{Index[]} set}
As we discussed before, only secret coefficients belonging to the secret set $S_{w}$ will still be undetected. Now, we will search the indexes of the secret coefficients that are still not decided and store them in a set named "\texttt{Index[]}". Then, we apply the parallel checking oracle $\mathcal{O}_{\mu}$ on this \texttt{Index[]} set. We describe the detailed process in the following section.

\subsubsection{Construction of ciphertext $c_{\alpha}$ from \texttt{Index[]}:}

Before arriving at this stage, we found most secrets except the \texttt{Index[]} set secrets. Without loss of generality, assume that $ \mathbf{s}[i]\in S_{w}$ $\forall i\in \texttt{Index[]}$, where $S_{w}$ contains the values with a low probability occurrence and $d_{w}$ is the corresponding value of ciphertext selection in the Figure \ref{generic}. 

Let $\texttt{Index[]}=\{\alpha_{0},\ \alpha_{1},\ldots,\ \alpha_{r}\}$.
Construct the dummy ciphertext $ct = ({\bf u},
{\bf v})\in R_{q}\times R_{q}$ to reduce the cardinality of the secret set corresponding to each coefficient of the secret coefficients $\mathbf{s}[\alpha_{0}],\ldots,\ \mathbf{s}[\alpha_{t-1}]$ (we called it secret block $\mathbf{s}_{\alpha_{0},\ldots,\ \alpha_{t-1}}^{(t)}$ of size $t$). We choose $\mathbf{u}[0] = k_{u}$ and $ \mathbf{v}[\alpha_{j}] = d_{w} $, $\forall 0\leq j\leq t-1$, as each ${\bf s}[\alpha_{j}]\in S_{w}$. All the remaining coefficients of ${\bf u}$ and ${\bf v}$ will be zero. 
Then the decrypted message will be 
\begin{equation}
m^j=
    \begin{cases}
        \mathtt{Decode}(d_{w} - k_{u}\mathbf{s}[j]) & \text{if } j=\alpha_{0},\ \ldots,\ \alpha_{t-1}\\\
        \mathtt{Decode}(-k_{u}\mathbf{s}[j]) & \text{Otherwise } 
    \end{cases}    
\end{equation}
So, the message will depend on all the $\alpha_{j}$-{th} secret coefficient $\mathbf{s}[\alpha_{j}]$, where $0\leq j\leq t-1$, which is followed by the construction of our binary decision tree shown in the Figure~\ref{generic}.

We query the oracle $\mathcal{O}_{\mu}$ with the forged ciphertext $c_{\alpha}$ and the $2^{t}$ messages
$m^{(0)'},\ m^{(1)'},\ldots,\ m^{(2^{t}-1)'}$ to get $t$ bits of information with location $\alpha_{0},\ldots,\ \alpha_{t-1}$ simultaneously. Here, we take each message $m^{(i)'}$ such that $\alpha_{j}$-{th} bit of the message $m^{(i)'}$ is the $j$-{th} bit of $i$ and the others bits are zero. Here, we use Algorithm~\ref{create_cipher_new} to create forged ciphertexts.

 \begin{algorithm}[t]
	\caption{Ciphertext creation II}\label{create_cipher_new}
 \hspace*{\algorithmicindent} \textbf{Input:} The index $\alpha_{0},\ldots,\ \alpha_{t-1}$ of those we want to find actual secret.\\
 \hspace*{\algorithmicindent}\textbf{Output:}  Ciphertext $ct$ such that the decrypted message of $c$ will be zero except the possitions $\alpha_{0},\ldots,\ \alpha_{t-1}$.
	\begin{algorithmic}[1]
            \For{$k=0; k<n; k++$}
                \State $\mathbf{u}[k]=0;\ \mathbf{v}[k]=0$
            \EndFor
            \State ${\bf u}[0]=k_{u}$
            \For{$k=0; k<t; k++$}
                \State $\mathbf{v}[\alpha_{k}]=d_{w}$
            \EndFor
\end{algorithmic}
\end{algorithm}
\subsubsection{Updating the secret coefficients whose index lies in \texttt{Index[]}:}
    
We divide the sampling set into two distinct parts: $S_{2w+1}=\{s: \mathtt{Decode}(d_{w} - k_{u}s)=0\}$ and $S_{2w+2}=\{s: \mathtt{Decode}(d_{w} - k_{u}s)=1\}$, where $d_{w}$ is a predefined constant. Since $S_{w}$ contains the values such that the highest distance from the root node with $|S_{w}|>1$, therefore $|S_{2w+1}|$ and $|S_{2w+2}|$ must be 1. Otherwise, it violates our assumption of the set $S_{w}$. 
So, querying the oracle $\mathcal{O}_{\mu}$ with one ciphertext $c_{\alpha}$ and the above messages $m^{(0)'},\ m^{(1)'},\ \dots,\ m^{(2^{t}-1)'}$, we will get a decrypted message as a response. This decrypted message decides the $t$ number of secret coefficients ${\bf s}[\alpha_{0}],\ {\bf s}[\alpha_{1}]\ \dots,\ {\bf s}[\alpha_{t-1}]$ at a time. 
So, running the process $\lceil\frac{\texttt{|Index[]|}}{t}\rceil$ times, we will find the whole secret with mixed signs and in a different order. We described the process of finding the secret in actual order. Also, from Equation \ref{gen_cipher}, we can see that for the secret block $s_{j'}^{(t)}$, each $j$-{th} message $m^{j}$ will depend on the secret coefficient $-s[j'+j]$, $0\leq j\leq t-1$. So basically, we are decreasing the dimension secret coefficients $\mathbf{s}[0],\ \ldots,\ \mathbf{s}[t-1],\ -\mathbf{s}[n-t],\ -\mathbf{s}[n-t+1]\dots,\ -\mathbf{s}[n-1],\ -\mathbf{s}[n-2t],\ldots,\ -\mathbf{s}[n-t-1],\ \dots-{\bf s}[t],\ -{\bf s}[t+1],\ \dots,\ -{\bf s}[2t-1]$. We transformed it into the actual secret block using the Algorithm~\ref{Rotr}. 
\begin{algorithm}[t]
	\caption{Rotating secret coefficients}\label{Rotr}
        \hspace*{\algorithmicindent} \textbf{Input:} The secret {\bf s} is in the sequence  $\mathbf{s}[0],\ \dots,\ \mathbf{s}[t-1],\ -\mathbf{s}[n-t],\ -\mathbf{s}[n-t+1],\ \ldots,\ -\mathbf{s}[n-1],\ -\mathbf{s}[n-2t],\ \dots,\ -\mathbf{s}[n-t-1],\ \dots={\bf s1}$\\
 \hspace*{\algorithmicindent}\textbf{Output:}  The secret ${\bf s}$ with actual order i.e.,$(\mathbf{s}[0],\ \mathbf{s}[1],\ \ldots,\ \mathbf{s}[n-1])$
	\begin{algorithmic}[1]
		\For{$j=0;j<t;j++$}
		\State $\mathbf{s}[j]=\mathbf{s1}[j];$
		\EndFor
		\For{$j=1;j<\lfloor \frac{n}{t}\rfloor;j++$}
		\For{$k=0;k<t;k++$}
		\State $\mathbf{s}[t*j+k]=-\mathbf{s1}[(n-t*j+k)\%n];$
		\EndFor
		\EndFor
		\State Return ${\bf s}$
	\end{algorithmic}
\end{algorithm}

\subsubsection{Number of queries:}
Here $(S_{w},\ d_{w})$ is the most distanced node from the root node with $|S_{w}|>1$ and containing secrets occurring with comparatively lower probability. 
\begin{enumerate}
	\item Best case: If all the secret values lie in $S_{0}-S_{w}$, then the number of queries will be minimum because, in this case, we need $\lfloor \log|S_{0}|\rfloor$ queries to find each block of secrets $\mathbf{s}_{i}[j],\ \mathbf{s}_{i}[j+1],\ \dots,\ \mathbf{s}_{i}[j+t-1]$ of blocksize $t$. The total number of queries will be: $\lceil \frac{n}{t}\rceil\times \lfloor \log|S_{0}|\rfloor$. 
	\item Average case: Let $E_{1}$ be the expected number the secret coefficients those belongs to $S_{w}.$ Then the total number of queries will be: $(\lceil \frac{n}{t}\rceil\times \lfloor \log|S_{0}|\rfloor)+(\lceil \frac{E_{1}}{t}\rceil)$. 
\end{enumerate} 
With our method, the number of queries for the average case decreases compared to the state-of-the-art works~\cite{ParallelPC,ParallelPC1}.
\subsection{Model for Kyber and Saber}
Kyber and Saber are based on the module-LWE and module-LWR problems, respectively i.e., here, the modules $R_{q}^{l}$ are used for the secret and the ciphertext $\Bar{\bf b}'$ instead of the ring $R_{q}$. But if we construct
$c=(\Bar{\bf b}',\ {\bf v})$ as follows: 
\begin{equation*}
		\Bar{\bf b}'_{i}[j]=
		\begin{cases}
			k_{u}\text{, if  } i=0,j=0  \\
			0  \text{,
   otherwise}
		\end{cases}
		\text{and }{\bf v}[j]=
		\begin{cases}
			k_{v_{j}}\text{, if  } 0\leq j\leq t-1 \\
			v \text{,
   otherwise}\,,
		\end{cases}
\end{equation*}
where $k_{u},k_{v_{j}}$ are constants. Then the problem reduces to the generic LPR problem, i.e., to the ring problem. Therefore, here the total number of queries will be $l\times \texttt{ the number of queries for LPR}$.
We use the corresponding $d_{i}$ from Table~\ref{Table01} and \ref{Table02} for Kyber768 and the Saber, respectively. We will construct the corresponding binary decision tree from Table~\ref{Table01} and \ref{Table02} and construct our ciphertext accordingly. For Kyber768, we have seen that for $k_{u}=38,\ v=14$ and $k_{v_{j}}=d_{i}$. From Table~\ref{Table01}, we can recover the secret by a similar process mentioned in the previous section.

\begin{table}[t]
\begin{minipage}{.5\linewidth}
\centering
\caption{For Kyber768}

\begin{tabular}{|p{.5cm}|l|l|l|l|}
\hline
  & \multicolumn{4}{|c|}{$u=38,v=14$}\\
\hline
 S& $d_{0}=12$ & $d_{1}=4$ & $d_{2}=13$ & $d_{4}=3$ \\ \hline
-2 & 0 & 1 & 0 & 1 \\ \hline
-1 & 0 & 1 & 0 & 0 \\ \hline
0 & 0 & 0 & 0 & 0 \\ \hline
1 & 1 & 0 & 0 & 0 \\ \hline
2 & 1 & 0 & 1 & 0 \\ \hline
\end{tabular}
\label{Table01}
\end{minipage}
\hspace{1cm}
\begin{minipage}{.45\linewidth}
\centering
\caption{ For Saber}

\begin{tabular}{|l|l|l|l|l|l|l|l|l|}
\hline
S  & \multicolumn{7}{|c|}{$u=0x3c8$}                                       & \multicolumn{1}{|l|}{$u=7$} \\
\hline
   & $d_{0}$ & $d_{2}$ & $d_{5}$ & $d_{6}$ & $d_{1}$ & $d_{3}$ & $d_{4}$ &  $d_{7}$            \\
   & $=4$ & $=2$ & $=3$ & $=1$ & $=6$ & $=7$ & $=5$ &  $=12$            \\
   \hline
-4 &  0       & 0        & 0        &     0    & 0        &   0      &  0       &  0                       \\
\hline
-3 &  0       &   0      & 0        &      0   &  0       &     1    &    0                &            0\\
\hline
-2 &  0       &  0       & 0        &       0  &   1      &   1      &   0                &        0    \\
\hline
-1 &  0       &   0      &  0       &       0  &   1      &   1      &   1                   &    0        \\
\hline
0  &   1      &   0      &  0       &        0 &    1     &  1       &   1                  &    0        \\
\hline
1  &  1       &   0      &  1       &         0& 1        &  1       &   1                   &    0        \\
\hline
2  &   1      &     1   &  1       &         0&      1   &   1      &    1                 &      0      \\
\hline
3  &   1      &    1     &  1       &  1       &  1       &  1       &    1                  &  0   \\
\hline
4  &   1      &    1     &   1      &    1     &    1     &    1     &      1                & 1    \\
\hline

\end{tabular}
\label{Table02}
\end{minipage}
\end{table}

\subsubsection{Number of queries for Kyber768 and Saber:}
According to Table~\ref{Table01}, for Kyber768 $S_{4}$ will be the highest distanced node from the root node containing secrets with comparatively low probability and $|S_{4}|>1$. Also, Table~\ref{Table02} shows that for Saber, $S_{7}$ will be that specified node above. For Kyber768 and Saber, $l=3$, we consider our best case and average cases of both the algorithms for $l=3$.   
\begin{enumerate}
\item Best case: In case of Kyber768, if all the secret values lie in $S_{0}-S_{4}$, then the number of queries will be minimum because, in this case, we need $2$ queries to find each block of secrets $\Bar{\mathbf{s}}_{i}[j],\ \Bar{\mathbf{s}}_{i}[j+1],\ \dots,\ \Bar{\mathbf{s}}_{i}[j+t-1]$ of blocksize $t$. The total number of queries will be: $\lceil \frac{n}{t}\rceil\times 3\times 2.$ 
For Saber this number will be $\lceil \frac{n}{t}\rceil\times 3\times 3.$

    \item Average case: In the case of Kyber768, if $E_{1}$ is the expected number of the secret coefficients of each polynomial that lie in the set $S_{4}$, the total number of queries will be: $3\times((\lceil \frac{n}{t}\rceil\times 2)+\lceil\frac{E_{1}}{t})\rceil).$ Similarly, for Saber, if $E_{1}$ is the expected number of the secret coefficient of each polynomial that lies in $S_{7}$, then the total number of queries will be: $3\times((\lceil \frac{n}{t}\rceil\times 3)+\lceil\frac{E_{1}}{t}\rceil).$
\end{enumerate} 

\subsection{Comparing our attack with the state-of-the-art}

In this section, we compare the total number of ciphertexts required to retrieve the whole secret key for the average case in Kyber768 and Saber with our attack and the work by Rajendran et al.~\cite{ParallelPC1}, which also proposed methods to reduce the number of ciphertexts using parallel plaintext checking oracle model. Even though we need to use the same number of ciphertext as~\cite{ParallelPC1} to recover the whole secret key when the parallelization factor $t=1$, our attack model requires less number of ciphertexts than \cite{ParallelPC1} to recover the whole secret key in the average case when the parallelization factor $t>1$. If $t=10$ or $12$ or $16$, for Kyber768 we use approximately $22\%$ less number of ciphertext than \cite{ParallelPC1}. Also, in Saber, if we take $t=10$, we require $\approx 39\%$ less number of ciphertext than the paper \cite{ParallelPC1} to recover the key. However, we require $57$ number of ciphertext to recover the whole secret key of Kyber768 in the average case when the parallelization factor $t=32$. We observe that increasing the parallelization factor $t$ will reduce the number of required ciphertexts. However, in this case, the process of finding the decrypted message from the shared key (offline calculation) will be more costly (takes ~$2^{t}$ comparison). For this reason, we take the value of the parallelization factor $t$ up to $32.$ But, with a more powerful computer that can do $2^{40}$ comparison, then we can take the parallelization factor $t=40$. In this case, the number of queries will be 48.

\subsubsection{Frequency of fault induction in the attack for Kyber768:}
We have discussed earlier that to recover the whole secret of the algorithm Kyber768, we require $57$ faulted shared keys i.e., $57$ many times, we often have to introduce the bit-flip faults at the location of the variable "fail".

\begin{table}[!ht]
\centering
\caption{Number of queries required to recover the key for Kyber768 and Saber in total}
\label{NumQuery}
\begin{tabular}{ll|lclclclclclc}
\hline
\multicolumn{2}{l|}{\multirow{2}{*}{Scheme}}                                                                                                                                   & \multicolumn{12}{c}{Parallelization factor t}                                                                                                                           \\ \cline{3-14} 
\multicolumn{2}{l|}{}                                                                                                                                                          &  & \multicolumn{1}{c|}{$1$}    &  & \multicolumn{1}{c|}{$10$}  &  & \multicolumn{1}{c|}{$12$}  &  & \multicolumn{1}{c|}{$16$}  &  & \multicolumn{1}{c|}{$32$} &  & $40$ \\ \hline
\multicolumn{1}{l|}{\multirow{2}{*}{Kyber768}} & \begin{tabular}[c]{@{}l@{}}This work\\ $3\times((\lceil \frac{256}{t}\rceil\times 2)+\lceil\frac{80}{t})\rceil)$\end{tabular} &  & \multicolumn{1}{c|}{$1776$} &  & \multicolumn{1}{c|}{{\bf 180}} &  & \multicolumn{1}{c|}{ \bf 153} &  & \multicolumn{1}{c|}{\bf 111} &  & \multicolumn{1}{c|}{$\bf 57$} &  & {$\bf 48$} \\ \cline{2-14} 
\multicolumn{1}{l|}{}                          & Rajendran et al.~\cite{ParallelPC1}                                                                                                            &  & \multicolumn{1}{c|}{$1776$} &  & \multicolumn{1}{c|}{$232$} &  & \multicolumn{1}{c|}{$197$} &  & \multicolumn{1}{c|}{$144$} &  & \multicolumn{1}{c|}{$72$} &  & $63$ \\ 
\hline
\multicolumn{1}{l|}{\multirow{2}{*}{Saber}}    & \begin{tabular}[c]{@{}l@{}}This work\\ $3\times((\lceil \frac{256}{t}\rceil\times 3)+\lceil\frac{9}{t})\rceil)$\end{tabular} &  & \multicolumn{1}{c|}{$2331$}       &  & \multicolumn{1}{c|}{$\bf 237$}      &  & \multicolumn{1}{c|}{$201$}      &  & \multicolumn{1}{c|}{$147$}      &  & \multicolumn{1}{c|}{$75$}     &  &   $66$  \\ \cline{2-14} 
\multicolumn{1}{l|}{}                          & Rajendran et al.~\cite{ParallelPC1}                                                                                                             &  & \multicolumn{1}{c|}{$-$}       &  & \multicolumn{1}{c|}{$390$}      &  & \multicolumn{1}{c|}{$-$}      &  & \multicolumn{1}{c|}{$-$}      &  & \multicolumn{1}{c|}{$-$}     &  & $-$     \\ 
\hline
\end{tabular}
\end{table}

\section{Realization of the fault model}
In this section, we are going to illustrate an end-to-end strategy to demonstrate the fault model in practice.
\subsection{Nature of the fault in the attack}

In the previous sections, we discuss that our objective is to obtain the output $\mathcal{F}(K',\ \mathcal{H}(ct))$ by exploiting a fault, where $K'$ is derived from the decrypted message $m$ of the ciphertext $ct$. This fault uses the plaintext checking oracle $\mathcal{O}_{\mu}$. To achieve this, it is crucial to neutralize the effectiveness of comparing two ciphertexts, denoted as $c$ and $ct$, in terms of equality checking.
For all security levels of Saber and Kyber, the design employs a {\tt verify} function that takes two ciphertexts, $c$ and $ct$, along with their lengths and returns $0$ if they are equal or $1$ otherwise. The result is stored in a variable called "fail". In our attack, we construct ciphertexts in a particular pattern, ensuring that the ciphertexts $c$ and $ct$ are highly likely to be unequal. As a result, the variable "fail" will always be set to $1$. This allows us to perform a bit-flip or get stuck at zero at the location of the "fail" variable, thus obtaining our desired output $\mathcal{F}(K',\ \mathcal{H}(ct))$. If we observe that for our constructed ciphertext $ct$, the value of the shared key is different from $\mathcal{F}(z,\ \mathcal{H}(ct))$. At this time, we are ensured that the value of the "fail" variable has changed to $0$, and this value is our essential shared key $\mathcal{F}(K',\ \mathcal{H}(ct))$.

 A stuck-at-zero fault is where a signal or a specific bit within a circuit is constantly held at logic zero. This fault can occur due to manufacturing defects, electrical shorts, environmental factors, or other physical issues. In contrast, a bit-flips fault involves the unintentional change of a single bit within a circuit or memory location from its intended value to the opposite value.
Both stuck-at-zero and bit-flip faults can have various causes and implications.
It is important to note that the specific type and cause of these faults can vary depending on the context, such as the hardware or software implementation, the cryptographic scheme used, and the fault injection techniques employed. Stuck-at-zero and bit-flip faults can lead to unexpected behaviour, data corruption, security vulnerabilities, or system crashes.
To ensure system reliability and data integrity, detecting and mitigating these faults often involves employing error detection and correction techniques, such as error-correcting codes, redundant storage methods, or fault-tolerant designs.

In this paper, we choose Dynamic Random Access Memory (DRAM) reliability issue {\it Rowhammer} to introduce a software-driven hardware fault attack to induce a bit-flip ($1\rightarrow 0$) at the address of the "fail" variables. We also present a series of steps that could be followed to incorporate this fault at a precise location in realistic timeframes.


\subsection{Our target devices}
To demonstrate our attack, we employ a deliberate technique of inducing bit-flips during the decapsulation process of Kyber. In our model, the attacker is assumed to be colocated in the same server as the victim, which performs the decapsulation process of Kyber and Saber. This scenario can also be extended to multiple virtual machines operating on a shared server. In this model, the primary assumption is that the victim and the attacker are co-located on the same physical piece of memory hardware, typically a DRAM and the vulnerable locations are neighbors to each other.
This model exists in the current research field of row hammer \cite{ExplFrame,FRODOFLIP} and is also consistent with most microarchitectural attacks \cite{rowFault}.
Furthermore, since Kyber and Saber are designed as a CCA-secure scheme, our attack assumes that the attacker can often query the decapsulation process with the constructed ciphertext. We demonstrate our attack against the machines listed in Table \ref{tab:my-table}.
\begin{table}[t]
\centering
\begin{tabular}{|l|l|l|}
\hline
   & Model name                                    & RAM size \\ \hline
1. & Intel (R) Core (TM) i7-4770 CPU & 4 GB     \\ \hline
2. & Intel (R) Core (TM) i7-3770 CPU & 8 GB     \\ \hline
3. & Intel (R) Core (TM) i5-3330 CPU & 4 GB     \\ \hline
\end{tabular}
\caption{Model details of our target devices}
\label{tab:my-table}
\end{table}
\subsection{Probabilities of incorporating precise fault using random Rowhammer}
The task of incorporating bit-flips in random locations in memory is common and is very well studied in literature after Rowhammer has been reported in practice, but the hard part is to precisely induce the faults in the location of one's choice. In this paper, considering the target example, if we run the target code of Kyber/Saber multiple times in one process and an unsupervised row hammer code in another process, the address of the variable "fail" coinciding with one of the vulnerable locations, the probability of such event occurrence is considerably low. Suppose there are a total $N$ number of vulnerable locations after hammering randomly among $N_{1}$ locations present on a device. Then,
\text{the possibility of the variable "fail" being vulnerable} = \text{Pr(the location of "fail"=X)} $\times$ \text{Pr("fail" coincide in a vulnerable location |the }\\ \text{location of "fail"=X)} $=\frac{1}{N_{1}} \times \frac{N}{N_{1}}$
$=\frac{N}{N^{2}_{1}}$, which is very low as $N_{1}\gg N$. 
In our system, we randomly access $N_{1} = 2^{30}$ bytes of memory; we discovered $N <10$ vulnerable locations by accessing the memory randomly. Notably, the number of vulnerable locations  ($N$) is considerably smaller than the total memory access. In order to make this process deterministic, we follow the steps described below.

\noindent
\textbf{Using the deterministic process of Rowhammer: }We have used the {\it hammertime} code\footnote{\url{"https://github.com/vusec/hammertime.git"}} available at~\cite{hammertime}   to execute row hammering operations. Through our exploration, we have observed that {\it hammertime} is a valuable simulator, offering a convenient approach to deterministically evaluate vulnerable locations. This versatile tool is purpose-built for testing, profiling, and simulating the Rowhammer DRAM attack, providing a comprehensive suite of capabilities for assessing the outcomes of exploits.

\begin{figure}[t]
	\centering
	\includegraphics[width=0.7\linewidth]{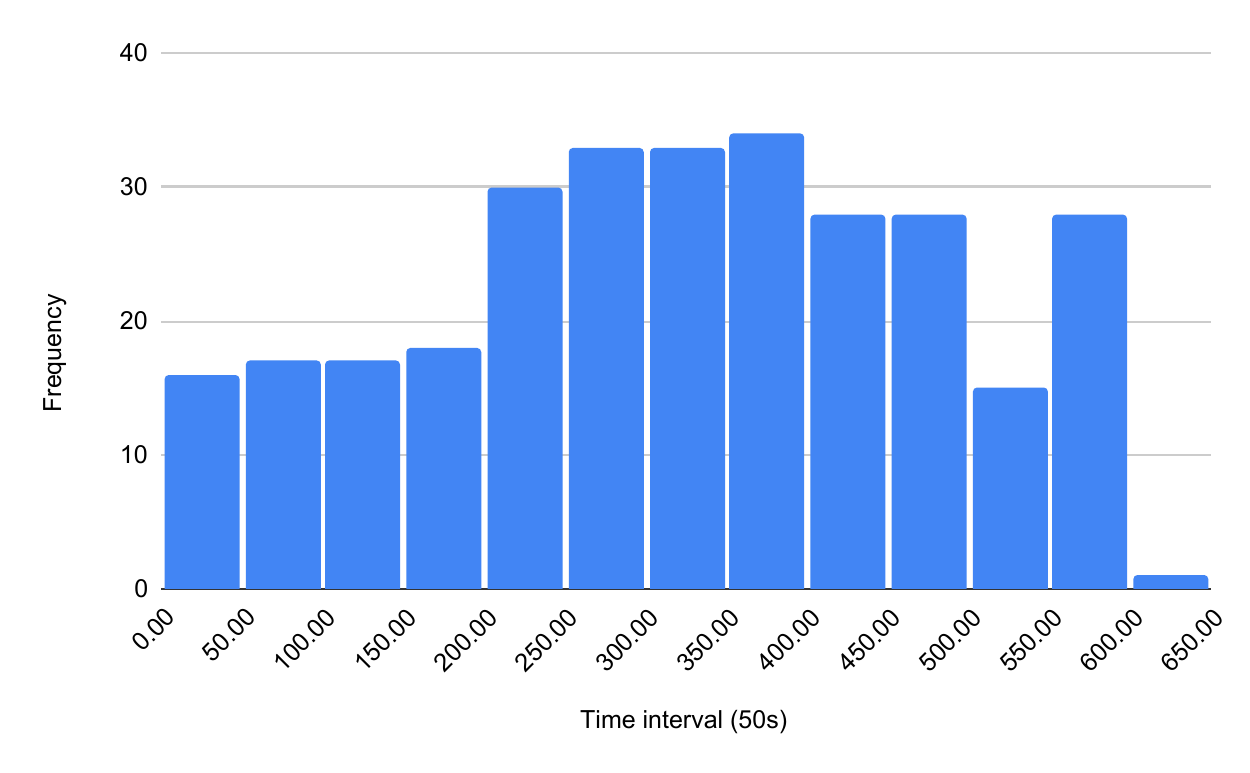}
	\caption{Frequency of bit-flips in every $50$ second}
	\label{hammeroffset}  
\end{figure}

\begin{figure}[t]
	\centering
	\includegraphics[width=0.7\linewidth]{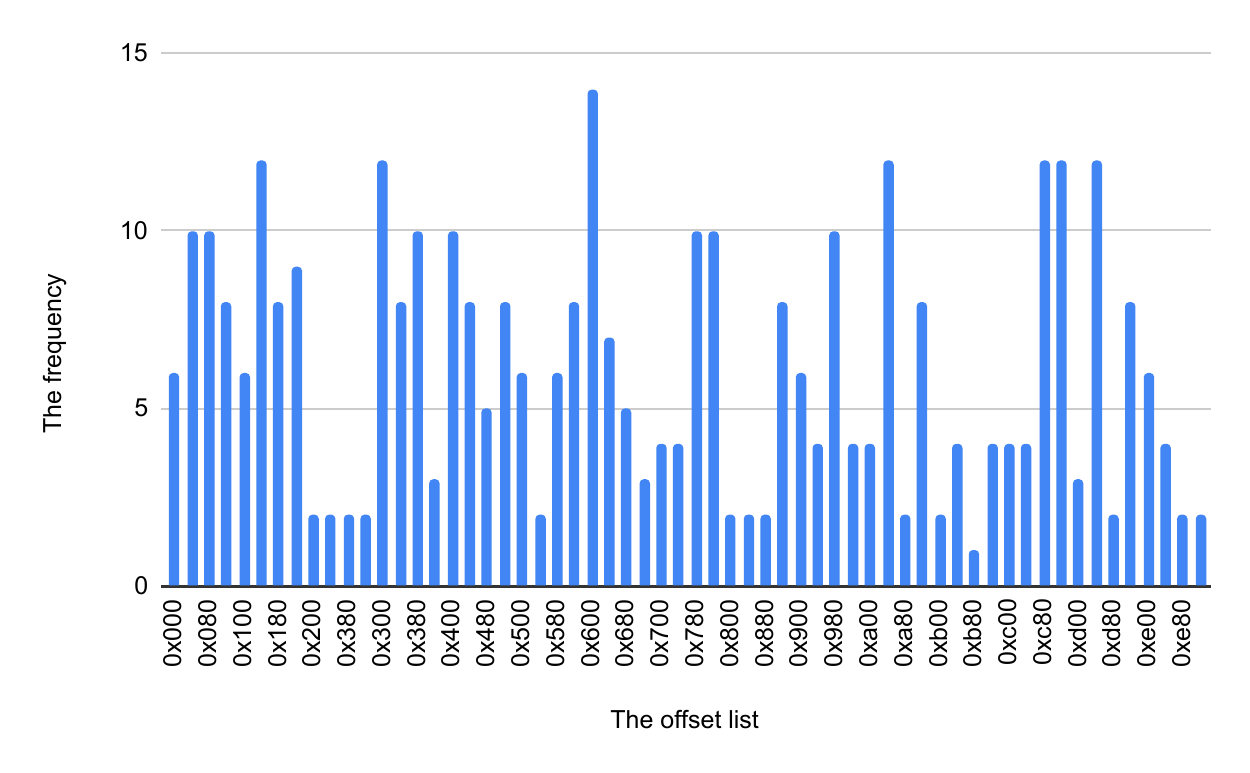}
 \caption{Frequency of the bit-flips in the corresponding offset.}
	\label{hammeroffset1}  
\end{figure}

\begin{figure}[b]
	\centering
	\scalebox{.38}{
		\begin{tikzpicture}
			\node[font=\Huge](a) at (-4,10){Process 1};
            \node[font=\Huge](b) at (10,10){Process 2};
            \node[font=\Huge](c) at (0,8){1. Run the "hammertime code" };
            
            \node[font=\Huge](d) at (0,7.1){until we get the bit-flip ($1\rightarrow 0$)};
            \node[font=\Huge](d) at (-2.5,6.2){at the location A};

            \node[font=\Huge](d) at (-1.5,4.8){2. Unmap the page of A};
            
            \node[font=\Huge](d) at (10.2,4.2){3.Run the decapsulation};
            \node[font=\Huge](d) at (10.2,3.3){ code of Kyber/Saber};
            
            \node[font=\Huge](f) at (-.1,3){4. Run the "hammertime code" };
            \node[font=\Huge](f) at (-.1,2.1){ for the same location A};
            
            \draw[font=\Huge][blue,-](5.7,10) -- (5.7,1);
             \draw[font=\Huge][red,-](-6,9) -- (5,9);
             \draw[font=\Huge][red,-](5.8,9) -- (15,9);
		\end{tikzpicture}
	}
	\caption{Template of generating oracle $\mathcal{O}_{\mu}$ using Rowhammer}
	\label{rowhammer_template}
\end{figure}
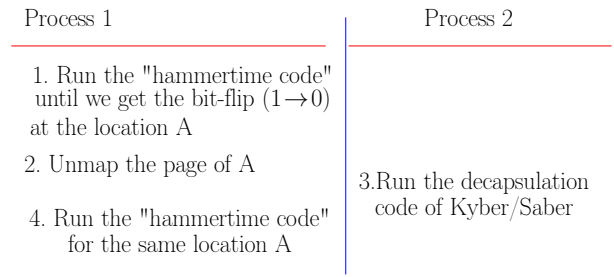

The provided code presents two types of row hammering techniques: single-sided row hammering and double-sided row hammering. We have employed the single-sided row hammering process outlined in their code for our implementation. In each iteration of this approach, our target is to find the vulnerable rows from an aggressive row's upper or lower rows. Also, in our victim machine, the bit-flip occurs considerably frequently. Figure~\ref{hammeroffset} shows the bit-flip frequency in every $50$ second.
In the hammertime code, we observe that this code deterministically selects an aggressive row, then fills up the memory with values all 1's ("$\tt{0xff}$") in the aggressive row and its neighbouring rows, and repeatedly flushes the corresponding portions of cache memory allocation. Iteratively, it only accesses the addresses with offsets $A=\{a_{i}\}_{i}$, where $a_{0}=0$ and $a_{i+1}-a_{i}=\tt{0x020}$, for all $i$ to check the bit-flip result. So, we can get the vulnerable address with the offset lie in $A$ by running the hammering code. Figure~\ref{hammeroffset1} shows the offsets of the bitflip addresses and their frequency observed in our experiments. 
We perform a first-level templating of main memory using the {\it hammertime} code as shown in Figure~\ref{hammeroffset1}, identifying locations that are vulnerable to Rowhammer. This templating step also aids us in identifying trigger rows so that we can replicate Rowhammer deterministically by re-accessing those aggressor rows again over time. 
 By using the hammering code, we get the vulnerable addresses having different offsets and construct the set $A$. In this particular attack algorithm, we want the adversary to induce a bit-flip to a known vulnerable location. In order to achieve that, the variable in the decapsulation process (target "fail" variable) must coincide with atleast one offset in the set $A$ of vulnerable addresses in order to precisely induce the fault. In order to increase the reproducibility of the attack over multiple runs, we have assigned the datatype of the variable "fail" in our implementation to "static int" rather than simply using "int". Doing so guarantees that the offset of the "fail" variable remains unchanged throughout the execution. Without loss of generality, if our attack methodology is implemented on any other target secret, then a similar technique could be applied to any global variable or a local variable with a static flag for the sake of the reproducibility of our attack.
We consider the offset $\tt{0x040}$ of the "fail" variable, which was observed on our executable. This offset can be any value without loss of generality in Kyber/ Saber's implementation, and the appropriate matching offset of the Rowhammer fault can also be selected from the templating phase. In our attack scenario, we select the vulnerable locations offset of $\tt{0x040}$ to show the vulnerability. We construct the following template shown in Figure~\ref{rowhammer_template}.

The templating method in Rowhammer provides a method that induces a bit-flip from $1$ to $0$ at the "fail" variable. First, we run the {\it hammertime} code and observe a bit-flip ($1\rightarrow 0$) at an address with the offset $\tt{0x040}$. In this phase, we proceed to unmap the corresponding page of that address and emit a signal, enabling us to execute the victim code in process 2. With a high likelihood, the victim code gets mapped to the unmapped page just being freed by the {\it hammertime} executable. This will allow the "fail" variable to be sitting in the Rowhammer vulnerable location of the unmapped page. The scenario of page reallocation of the recently unmapped page is commonly encountered using the Page Frame Cache during page allocations involving the buddy allocator \cite{ExplFrame}. 

 After successfully aligning the "fail" variable with the vulnerable location of Rowhammer, our objective is to actually induce the fault in the target location to change its value to "0". To accomplish this, we need to continue performing row hammering on the same aggressive row that inflicted the Rowhammer in the templating phase. This ensures that the bit-flip occurs at the same vulnerable address, which is now unmapped from the hammering code, but possessed by the target executable of the victim. To achieve this, we made some modifications to the {\it hammertime} tool, and iterated through the following processes. 
 
 An extra {\it loop} is added inside the {\it profile\_singlesided} function. Once the target page is unmapped, only then this {\it loop} will run. The {\it loop} contains minor modifications to the following functions {\it fill\_rows} and {\it c->hamfunc}. This modification involves a checking condition that inside the function {\it fill\_rows}, we ignore the addresses lying on the unmapped page. This function activates aggressive rows and neighbor rows and as a result, the vulnerable address is affected, leading to a change in its bit from "1" to "0".

After unmapping the page, we run the victim code (decapsulation process with our constructed ciphertext) parallel to the {\it hammertime code} until we observe the faulty shared key. If we observe a different shared key, then the Rowhammer attempt has been successful and we stop this process. 
We summarize the whole process as follows:
\begin{enumerate}
    \item By running the hammering code, vulnerable addresses with offsets from set A are identified, and the "fail" variable is positioned to coincide with one of these vulnerable addresses. A suitable vulnerable location is selected and the corresponding page is unmapped from the code.
    \item After unmapping the page, we run the victim code until we do not get the faulty shared key. If we get a different shared key, then we are done. 
    \item To achieve a bit-flip from "1" to "0" at the "fail" variable, row hammering is continued on the same aggressive row, modifying the fill\_row function to fill memory with "$\tt{0xff}$" and performing a memory flush on all addresses except the unmapped page corresponding to the vulnerable address.
\end{enumerate}

\begin{figure}[t]
	\centering
	\includegraphics[width=0.7\linewidth]{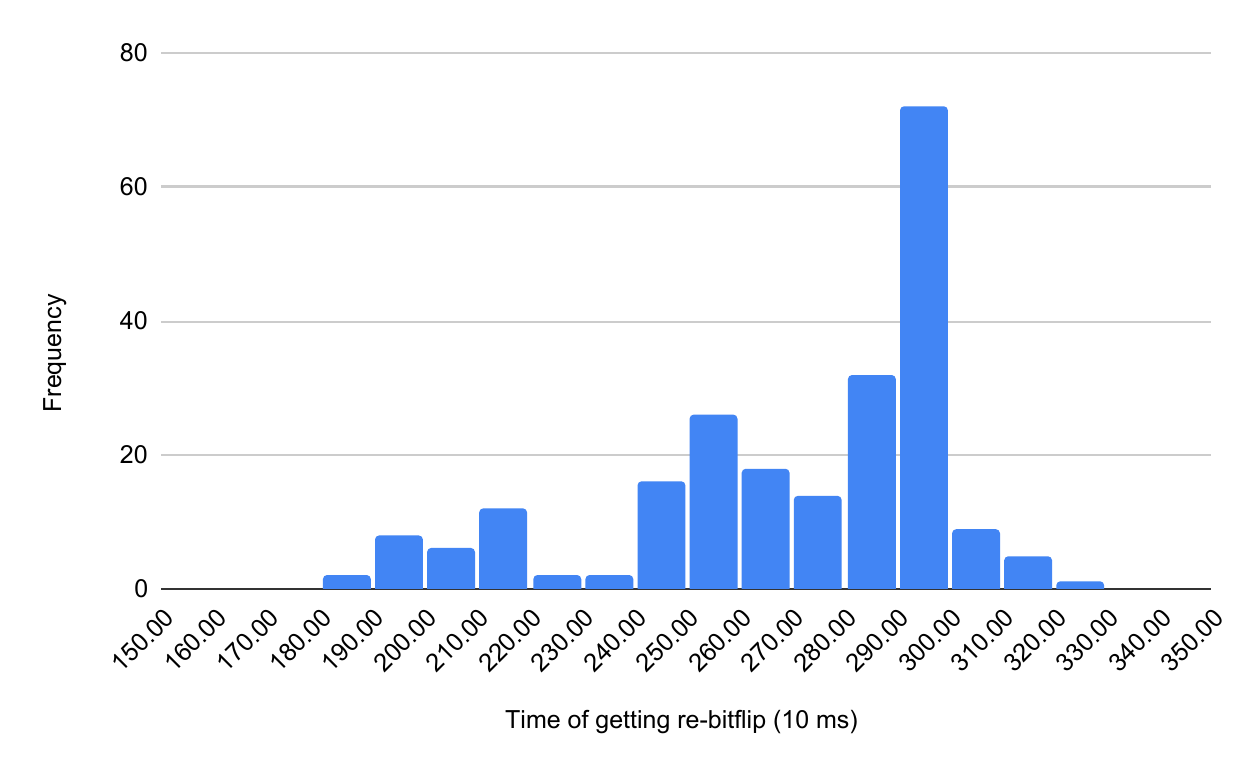}
	\caption{The value of an interval $[a,b]$ is the number of bit-flips which takes the time $t\in[a,b]$, to make re-bitflip at the same address }
	\label{Rebitflip}  
\end{figure}

Figure~\ref{Rebitflip} illustrates the distribution of timings observed for the Rowhammer bit-flip to occur at the vulnerable location through the {\it hammertime} code after unmapping the vulnerable page.
In order to estimate the total time to recover the whole secret key we need $57$ independent queries to the oracle. This translates to $57$ independent fault occurrences on the "fail" variable in the implementation of the decapsulation algorithm. One such occurrence can be estimated to happen in $< 350ms$ with a significantly high probability. So this attack can be realised using an additive progression of timing on respective queries and can be observed in a linear timescale.

\section{Discussion and future direction}
In this paper, we show an end-to-end software-driven hardware fault on PQ LWE-based KEMs. We choose Saber and Kyber key encapsulation schemes and perform the fault analysis with as much as $39\%$ reduced number of queries for Saber and approximately $23\%$ for Kyber768 on the existing literature. This was achieved by pruning selected leaves of the decisional binary search tree used in the attack. The fault induction using the Rowhammer has been known in the literature to appear in random locations of memory due to the reliability issues of commercial DDR RAMs. We follow some precise steps by first templating the memory space, listing out vulnerable addresses of a system, and then precisely locating the target KEM implementation in that vulnerable location. In this context, we use publicly available {\it Hammertime} code to template the memory space, then make minor modifications to re-induce Rowhammer using the selected aggressor rows on that same location deterministically. This semi-deterministic process is highly useful in conjunction with the paging policies of the Buddy allocator, and then inflicting these bit-flips on the publicly available target implementation. 

Though there has been recent work on Frodo KEM~\cite{FRODOFLIP}, where the authors incorporate fault in the key generation phase using Rowhammer. As discussed in Section~\ref{sec:attack_surface}, the key generation of a CCA-secure KEM is a one-time operation and is invoked rarely. Hence, if necessary the key generation can even be done offline in an isolated environment. On the other hand, the decapsulation of a CCA-secure KEM is invoked multiple times to generate the shared secret key from multiple sources. Therefore, in a practical scenario for the sake of performance, the decapsulation cannot run in an isolated environment. Therefore the attack described in~\cite{FRODOFLIP} is far less realistic than our attack methodology. 
Further, the authors assume that they can slow down the execution by slowing down components of the target executable. This is already a strong assumption. Additionally, the authors have disabled ASLR (Address Space Layout Randomization) for their experiments which makes the assumptions even stronger and the attack more unrealistic. 

\subsection{Shuffling and Masking:} Previous attacks~\cite{ParallelPC,ParallelPC1} based on parallel plaintext checking oracle have used side-channel analysis such as EM power analysis. So, these attacks can be prevented using masking countermeasures~\cite{DBLP:journals/iacr/RaviCB22}. Our attack can be conducted on the masked or shuffled implementation of the LWE-based KEMs. Because here, we do not use any side-channel assistance to perform the attack. We induce a bitflip fault to the "fail" variable, which stores the result of the comparison between the public ciphertext and the re-encrypted ciphertext. As a result of this fault, the value of the "fail" variable always remains 0, and that causes decapsulation success. When applying side-channel countermeasures such as masking and shuffling on the decapsulation algorithm of LWE-based KEMs~\cite{Masked_kyber,proof_shuffle,HigherMasking}, this fail variable remains unaffected and unmasked, since it is not dependent on the secret. Therefore, the success of our attack does not get affected by generic side-channel countermeasures such as masking or shuffling.

\subsection{Extension of our attack on other PQC schemes:} The parallel plaintext checking oracle used in our attack model can be applicable to any LWE-based KEMs. It is not specific to Saber and Kyber. It can be applicable to other LWE-based schemes such as NewHope~\cite{AlkimDPS16}, Lizard~cite{CheonKLS18}, Round5~\cite{BaanBFGLRSTZ19}, Frodo~\cite{frodo_scheme}, Smaug~\cite{cryptoeprint:2023/739}(proposed in the ongoing Korean PQC~\cite{web:kpqc} competition), etc. The Rowhammer methodology we propose in this work to introduce fault can also be applicable to other fault attack models where a single or multi-bit fault is required. Popular side-channel countermeasures such as masking and shuffling are ineffective to protect against this attack.

\subsection{Combining of lattice reduction techniques with our attack:} There can be some cases when the attacker only has a limited number of accesses to the decapsulation procedure. Then, the attacker can use our attack to recover some of the coefficients of the secret key and then use lattice reduction techniques to recover the rest of the secret key~\cite{lattice_reduction}. The LWE-estimator toolbox~\cite{LWE_estimator,leaky_estimator} can provide an estimate on the computation effort required to recover the secrets using the lattice reduction techniques. It is up to the attacker to determine the optimum point till when our attack should be stopped and the lattice reduction methods should be used. However, more investigation is needed to combine our attack results with these LWE-estimators to efficiently recover the secret key. We would like to investigate it in the future.


\subsection{Possible countermeasures} 
Although masking or shuffling countermeasures are unable to prevent our attack, there are a few countermeasures that can be useful to thwart our attack. Below, we list these countermeasures in two categories.
\begin{itemize}
    \item Fault attack countermeasure on the LWE-based schemes:
Recently, Berthet et al.~\cite{countermeasure1} propose a countermeasure named quasi-linear masking on Kyber to prevent fault injection attacks together with side-channel attacks. This countermeasure might be used to prevent our attack.
    \item Rowhammer Countermeasures:
There have been various countermeasures of RowHammer attacks proposed in the literature. The authors in the paper~\cite{kim2014flipping} proposed Probabilistic Adjacent Row Activation (PARA), where the memory controller is designed to refresh its adjacent rows with probability p (typically 1/2). The memory controller being probabilistic, the approach does not require a complex data structure for counting the number of row activations. Earlier in~\cite{seaborn2015exploiting}, it was shown that doubling the refresh rate and removing access to clflush instruction are potential prevention techniques to RowHammer. An interesting countermeasure to rowhammer has been proposed in Anvil~\cite{aweke2016anvil}. If the cache misses over a time interval is observed to be significantly high, then the software module triggers sampling of the DRAM accesses. ANVIL selectively performs a row refresh if the software module detects repeated accesses to particular rows in the same bank.
Another process, Target Row Refresh (TRR), believed to be a definitive solution, can prevent RowHammer bit flips~\cite{Datasheet_ddr4} [1]. However, in the paper~\cite{TRResspass}, the authors also find that consumer CPUs rely on in-DRAM TRR and are vulnerable to many-sided RowHammer attacks. They introduce TRRespass, which can autonomously discover intricate hammering patterns to launch real-world attacks on numerous DDR4 DRAM modules available in the market. Till now, there is no concrete solution that can prevent the RowHammer bit flip problem.
[1] J.-B. Lee, “Green Memory Solution,” in Samsung Electronics, Investor’s Forum, 2014.
\end{itemize}

\section*{Acknowledgements}

This work was supported in part by Horizon 2020 ERC Advanced Grant (101020005 Belfort), CyberSecurity Research Flanders with reference number VR20192203, BE QCI: Belgian-QCI (3E230370) (see beqci.eu), and Intel Corporation. 

Angshuman Karmakar is funded by FWO (Research Foundation – Flanders) as a junior post-doctoral fellow (contract number 203056 / 1241722N LV). Puja Mondal and Angshuman Karmakar are also supported by C3iHub, IIT Kanpur.

%
%
%
 \bibliographystyle{splncs04}
 \bibliography{main}

\begin{thebibliography}{10}
\providecommand{\url}[1]{\texttt{#1}}
\providecommand{\urlprefix}{URL }
\providecommand{\doi}[1]{https://doi.org/#1}

\bibitem{nist_final_report}
Alagic, G., Apon, D., Cooper, D., Dang, Q., Dang, T., Kelsey, J., Lichtinger, J., Liu, Y.K., Miller, C., Moody, D., Peralta, R., Perlner, R., Robinson, A., Smith-Tone, D.: {Status Report on the Third Round of the NIST Post-Quantum Cryptography Standardization Process}. Online. Accessed 26th June, 2023 (2022), \url{https://nvlpubs.nist.gov/nistpubs/ir/2022/NIST.IR.8413-upd1.pdf}

\bibitem{LWE_estimator}
Albrecht, M.R., Player, R., Scott, S.: {On the concrete hardness of Learning with Errors}. Cryptology ePrint Archive, Report 2015/046 (2015), \url{https://eprint.iacr.org/2015/046}

\bibitem{AlkimDPS16}
Alkim, E., Ducas, L., P{\"{o}}ppelmann, T., Schwabe, P.: {Post-quantum Key Exchange - {A} New Hope}. In: Holz, T., Savage, S. (eds.) 25th {USENIX} Security Symposium, {USENIX} Security 16, Austin, TX, USA, August 10-12, 2016. pp. 327--343. {USENIX} Association (2016), \url{https://www.usenix.org/conference/usenixsecurity16/technical-sessions/presentation/alkim}

\bibitem{aranha_power_attack}
Aranha, D.F., Fouque, P.A., G{\'e}rard, B., Kammerer, J.G., Tibouchi, M., Zapalowicz, J.C.: {GLV/GLS Decomposition, Power Analysis, and Attacks on ECDSA Signatures with Single-Bit Nonce Bias}. In: Sarkar, P., Iwata, T. (eds.) Advances in Cryptology -- ASIACRYPT 2014. pp. 262--281. Springer Berlin Heidelberg, Berlin, Heidelberg (2014)

\bibitem{ladderleak}
Aranha, D.F., Novaes, F.R., Takahashi, A., Tibouchi, M., Yarom, Y.: {LadderLeak: Breaking ECDSA with Less than One Bit of Nonce Leakage}. In: Proceedings of the 2020 ACM SIGSAC Conference on Computer and Communications Security. p. 225–242. CCS '20, Association for Computing Machinery, New York, NY, USA (2020). \doi{10.1145/3372297.3417268}, \url{https://doi.org/10.1145/3372297.3417268}

\bibitem{web:sphincs}
Aumasson, J.P., Bernstein, D.J., Beullens, W., Dobraunig, C., Eichlseder, M., Fluhrer, S., Gazdag, S.L., Hülsing, A., Kampanakis, P., Kölbl, S., Lange, T., Lauridsen, M.M., Mendel, F., Niederhagen, R., Rechberger, C., Rijneveld, J., Schwabe, P., Westerbaan, B.: {SPHINCS+: Stateless hash-based signatures}, \url{https://sphincs.org/}, [Online; accessed 28-June-2023]

\bibitem{FA_RSA}
Aum{\"u}ller, C., Bier, P., Fischer, W., Hofreiter, P., Seifert, J.P.: {Fault Attacks on RSA with CRT: Concrete Results and Practical Countermeasures}. In: Kaliski, B.S., Ko{\c{c}}, {\c{c}}.K., Paar, C. (eds.) Cryptographic Hardware and Embedded Systems - CHES 2002. pp. 260--275. Springer Berlin Heidelberg, Berlin, Heidelberg (2003)

\bibitem{aweke2016anvil}
Aweke, Z.B., Yitbarek, S.F., Qiao, R., Das, R., Hicks, M., Oren, Y., Austin, T.: {ANVIL: Software-based protection against next-generation rowhammer attacks}. ACM SIGPLAN Notices  \textbf{51}(4),  743--755 (2016)

\bibitem{BaanBFGLRSTZ19}
Baan, H., Bhattacharya, S., Fluhrer, S.R., Garc{\'{\i}}a{-}Morch{\'{o}}n, {\'{O}}., Laarhoven, T., Rietman, R., Saarinen, M.O., Tolhuizen, L., Zhang, Z.: {Round5: Compact and Fast Post-quantum Public-Key Encryption}. In: Ding, J., Steinwandt, R. (eds.) Post-Quantum Cryptography - 10th International Conference, PQCrypto 2019, Chongqing, China, May 8-10, 2019 Revised Selected Papers. Lecture Notes in Computer Science, vol. 11505, pp. 83--102. Springer (2019). \doi{10.1007/978-3-030-25510-7\_5}, \url{https://doi.org/10.1007/978-3-030-25510-7\_5}

\bibitem{BanerjeePR12}
Banerjee, A., Peikert, C., Rosen, A.: {Pseudorandom Functions and Lattices}. In: Pointcheval, D., Johansson, T. (eds.) Advances in Cryptology - {EUROCRYPT} 2012 - 31st Annual International Conference on the Theory and Applications of Cryptographic Techniques, Cambridge, UK, April 15-19, 2012. Proceedings. Lecture Notes in Computer Science, vol.~7237, pp. 719--737. Springer (2012). \doi{10.1007/978-3-642-29011-4\_42}, \url{https://doi.org/10.1007/978-3-642-29011-4\_42}

\bibitem{countermeasure1}
Berthet, P., Tavernier, C., Danger, J., Sauvage, L.: {Quasi-linear Masking to Protect Kyber against both {SCA} and {FIA}}. {IACR} Cryptol. ePrint Arch. p.~1220 (2023), \url{https://eprint.iacr.org/2023/1220}

\bibitem{DFA_EC}
Biehl, I., Meyer, B., M{\"u}ller, V.: {Differential Fault Attacks on Elliptic Curve Cryptosystems}. In: Bellare, M. (ed.) Advances in Cryptology --- CRYPTO 2000. pp. 131--146. Springer Berlin Heidelberg, Berlin, Heidelberg (2000)

\bibitem{frodo_scheme}
Bos, J.W., Costello, C., Ducas, L., Mironov, I., Naehrig, M., Nikolaenko, V., Raghunathan, A., Stebila, D.: {Frodo: Take off the Ring! Practical, Quantum-Secure Key Exchange from {LWE}}. In: Weippl, E.R., Katzenbeisser, S., Kruegel, C., Myers, A.C., Halevi, S. (eds.) Proceedings of the 2016 {ACM} {SIGSAC} Conference on Computer and Communications Security, Vienna, Austria, October 24-28, 2016. pp. 1006--1018. {ACM} (2016). \doi{10.1145/2976749.2978425}, \url{https://doi.org/10.1145/2976749.2978425}

\bibitem{KYBER}
Bos, J.W., Ducas, L., Kiltz, E., Lepoint, T., Lyubashevsky, V., Schanck, J.M., Schwabe, P., Stehl{\'{e}}, D.: {{CRYSTALS} - {Kyber: a CCA-secure module-lattice-based KEM}} (2017), \url{http://eprint.iacr.org/2017/634}

\bibitem{Masked_kyber}
Bos, J.W., Gourjon, M., Renes, J., Schneider, T., van Vredendaal, C.: {Masking Kyber: First- and Higher-Order Implementations}. {IACR} Trans. Cryptogr. Hardw. Embed. Syst.  \textbf{2021}(4),  173--214 (2021). \doi{10.46586/tches.v2021.i4.173-214}, \url{https://doi.org/10.46586/tches.v2021.i4.173-214}

\bibitem{ExplFrame}
Chakraborty, A., Bhattacharya, S., Saha, S., Mukhopadhyay, D.: {ExplFrame: Exploiting Page Frame Cache for Fault Analysis of Block Ciphers}. In: 2020 Design, Automation {\&} Test in Europe Conference {\&} Exhibition, {DATE} 2020, Grenoble, France, March 9-13, 2020. pp. 1303--1306. {IEEE} (2020). \doi{10.23919/DATE48585.2020.9116219}, \url{https://doi.org/10.23919/DATE48585.2020.9116219}

\bibitem{rowFault}
Chakraborty, A., Bhattacharya, S., Saha, S., Mukhopdhyay, D.: {Rowhammer Induced Intermittent Fault Attack on ECC-hardened memory} (2020), \url{https://eprint.iacr.org/2020/380}

\bibitem{cryptoeprint:2023/739}
Cheon, J.H., Choe, H., Hong, D., Yi, M.: {SMAUG: Pushing Lattice-based Key Encapsulation Mechanisms to the Limits}. Cryptology ePrint Archive, Paper 2023/739 (2023), \url{https://eprint.iacr.org/2023/739}, \url{https://eprint.iacr.org/2023/739}

\bibitem{leaky_estimator}
Dachman-Soled, D., Ducas, L., Gong, H., Rossi, M.: {{LWE} with Side Information: Attacks and Concrete Security Estimation}. Cryptology ePrint Archive, Report 2020/292 (2020), \url{https://eprint.iacr.org/2020/292}

\bibitem{AES}
Daemen, J., Rijmen, V.: {Rijndael for AES}. In: The Third Advanced Encryption Standard Candidate Conference, April 13-14, 2000, New York, New York, {USA}. pp. 343--348. National Institute of Standards and Technology, (2000)

\bibitem{SABER}
D'Anvers, J., Karmakar, A., Roy, S.S., Vercauteren, F.: {{Saber: Module-LWR based key exchange, CPA-secure encryption and CCA-secure} {KEM}} (2018), \url{http://eprint.iacr.org/2018/230}

\bibitem{DILITHIUM}
Ducas, L., Lepoint, T., Lyubashevsky, V., Schwabe, P., Seiler, G., Stehl{\'{e}}, D.: {{CRYSTALS} - {Dilithium: Digital Signatures from Module Lattices}} (2017), \url{http://eprint.iacr.org/2017/633}

\bibitem{FRODOFLIP}
Fahr, M., Kippen, H., Kwong, A., Dang, T., Lichtinger, J., Dachman-Soled, D., Genkin, D., Nelson, A., Perlner, R., Yerukhimovich, A., Apon, D.: When frodo flips: End-to-end key recovery on frodokem via rowhammer. In: Proceedings of the 2022 ACM SIGSAC Conference on Computer and Communications Security. p. 979–993. CCS '22, Association for Computing Machinery, New York, NY, USA (2022). \doi{10.1145/3548606.3560673}, \url{https://doi.org/10.1145/3548606.3560673}

\bibitem{CacheAttack1}
Fan, H., Wang, W., Wang, Y.: {Cache attack on {MISTY1}}. {IACR} Cryptol. ePrint Arch. p.~723 (2021), \url{https://eprint.iacr.org/2021/723}

\bibitem{web:falcon}
Fouque, P.A., Hoffstein, J., Kirchner, P., Lyubashevsky, V., Pornin, T., Prest, T., Ricosset, T., Seiler, G., Whyte, W., Zhang, Z.: {Falcon: Fast-Fourier Lattice-based Compact Signatures over NTRU} (2018), \url{https://falcon-sign.info/falcon.pdf}, [Online; accessed 28-June-2023]

\bibitem{TRResspass}
Frigo, P., Vannacci, E., Hassan, H., van~der Veen, V., Mutlu, O., Giuffrida, C., Bos, H., Razavi, K.: {TRRespass: Exploiting the Many Sides of Target Row Refresh}. In: 2020 {IEEE} Symposium on Security and Privacy, {SP} 2020, San Francisco, CA, USA, May 18-21, 2020. pp. 747--762. {IEEE} (2020). \doi{10.1109/SP40000.2020.00090}, \url{https://doi.org/10.1109/SP40000.2020.00090}

\bibitem{FOT}
Fujisaki, E., Okamoto, T.: {Secure Integration of Asymmetric and Symmetric Encryption Schemes}. J. Cryptol.  \textbf{26}(1),  80--101 (2013). \doi{10.1007/s00145-011-9114-1}, \url{https://doi.org/10.1007/s00145-011-9114-1}

\bibitem{lattice_reduction}
Gama, N., Nguyen, P.Q.: Predicting lattice reduction. In: Smart, N.P. (ed.) Advances in Cryptology - {EUROCRYPT} 2008, 27th Annual International Conference on the Theory and Applications of Cryptographic Techniques, Istanbul, Turkey, April 13-17, 2008. Proceedings. Lecture Notes in Computer Science, vol.~4965, pp. 31--51. Springer (2008). \doi{10.1007/978-3-540-78967-3\_3}, \url{https://doi.org/10.1007/978-3-540-78967-3\_3}

\bibitem{TimingPQC}
Guo, Q., Johansson, T., Nilsson, A.: {A Key-Recovery Timing Attack on Post-quantum Primitives Using the Fujisaki-Okamoto Transformation and Its Application on FrodoKEM}. In: Micciancio, D., Ristenpart, T. (eds.) Advances in Cryptology - {CRYPTO} 2020 - 40th Annual International Cryptology Conference, {CRYPTO} 2020, Santa Barbara, CA, USA, August 17-21, 2020, Proceedings, Part {II}. Lecture Notes in Computer Science, vol. 12171, pp. 359--386. Springer (2020). \doi{10.1007/978-3-030-56880-1\_13}

\bibitem{HermelinkPP21}
Hermelink, J., Pessl, P., P{\"{o}}ppelmann, T.: {Fault-Enabled Chosen-Ciphertext Attacks on Kyber}. In: Adhikari, A., K{\"{u}}sters, R., Preneel, B. (eds.) Progress in Cryptology - {INDOCRYPT} 2021 - 22nd International Conference on Cryptology in India, Jaipur, India, December 12-15, 2021, Proceedings. Lecture Notes in Computer Science, vol. 13143, pp. 311--334. Springer (2021). \doi{10.1007/978-3-030-92518-5\_15}

\bibitem{DilithiumCore}
Islam, S., Mus, K., Singh, R., Schaumont, P., Sunar, B.: {Signature Correction Attack on Dilithium Signature Scheme}. In: 7th {IEEE} European Symposium on Security and Privacy, EuroS{\&}P 2022, Genoa, Italy, June 6-10, 2022. pp. 647--663. {IEEE} (2022). \doi{10.1109/EuroSP53844.2022.00046}, \url{https://doi.org/10.1109/EuroSP53844.2022.00046}

\bibitem{Jiang2017}
Jiang, H., Zhang, Z., Chen, L., Wang, H., Ma, Z.: {Post-quantum IND-CCA-secure KEM without Additional Hash}. Cryptology ePrint Archive, Report 2017/1096 (2017), \url{https://eprint.iacr.org/2017/1096}

\bibitem{kim2014flipping}
Kim, Y., Daly, R., Kim, J., Fallin, C., Lee, J.H., Lee, D., Wilkerson, C., Lai, K., Mutlu, O.: {Flipping bits in memory without accessing them: An experimental study of DRAM disturbance errors}. ACM SIGARCH Computer Architecture News  \textbf{42}(3),  361--372 (2014)

\bibitem{web:kpqc}
KpqC: {Korean post-quantum cryptography competition} (2022), \url{https://www.kpqc.or.kr/competition.html}, [Online; accessed 28-June-2023]

\bibitem{HigherMasking}
Kundu, S., D'Anvers, J., Beirendonck, M.V., Karmakar, A., Verbauwhede, I.: {Higher-Order Masked Saber}. In: Galdi, C., Jarecki, S. (eds.) Security and Cryptography for Networks - 13th International Conference, {SCN} 2022, Amalfi, Italy, September 12-14, 2022, Proceedings. Lecture Notes in Computer Science, vol. 13409, pp. 93--116. Springer (2022). \doi{10.1007/978-3-031-14791-3\_5}, \url{https://doi.org/10.1007/978-3-031-14791-3\_5}

\bibitem{ReadBit}
Kwong, A., Genkin, D., Gruss, D., Yarom, Y.: Rambleed: Reading bits in memory without accessing them (05 2020). \doi{10.1109/SP40000.2020.00020}

\bibitem{MLWE}
Langlois, A., Stehl{\'{e}}, D.: {Worst-case to average-case reductions for module lattices}. Des. Codes Cryptogr.  \textbf{75}(3),  565--599 (2015). \doi{10.1007/s10623-014-9938-4}, \url{https://doi.org/10.1007/s10623-014-9938-4}

\bibitem{LPR}
Lyubashevsky, V., Peikert, C., Regev, O.: {On Ideal Lattices and Learning with Errors over Rings}. In: Gilbert, H. (ed.) Advances in Cryptology - {EUROCRYPT} 2010, 29th Annual International Conference on the Theory and Applications of Cryptographic Techniques, Monaco / French Riviera, May 30 - June 3, 2010. Proceedings. Lecture Notes in Computer Science, vol.~6110, pp. 1--23. Springer (2010). \doi{10.1007/978-3-642-13190-5\_1}, \url{https://doi.org/10.1007/978-3-642-13190-5\_1}

\bibitem{Datasheet_ddr4}
Micron: {DDR4 SDRAM Datasheet} (2016)

\bibitem{ECC_miller_Crypto86}
Miller, V.S.: {Use of Elliptic Curves in Cryptography}. In: Williams, H.C. (ed.) Advances in Cryptology --- CRYPTO '85 Proceedings. pp. 417--426. Springer Berlin Heidelberg, Berlin, Heidelberg (1986)

\bibitem{DBLP:journals/iacr/MujdeiBBKWV22}
Mujdei, C., Beckers, A., Bermundo, J., Karmakar, A., Wouters, L., Verbauwhede, I.: {Side-Channel Analysis of Lattice-Based Post-Quantum Cryptography: Exploiting Polynomial Multiplication}. {IACR} Cryptol. ePrint Arch. p.~474 (2022), \url{https://eprint.iacr.org/2022/474}

\bibitem{LUOV}
Mus, K., Islam, S., Sunar, B.: {QuantumHammer: A Practical Hybrid Attack on the LUOV Signature Scheme}. In: Proceedings of the 2020 ACM SIGSAC Conference on Computer and Communications Security. p. 1071–1084. CCS '20, Association for Computing Machinery, New York, NY, USA (2020). \doi{10.1145/3372297.3417272}, \url{https://doi.org/10.1145/3372297.3417272}

\bibitem{Flippingbit1}
Mutlu, O., Kim, J.S.: {RowHammer: {A} Retrospective}. {IEEE} Trans. Comput. Aided Des. Integr. Circuits Syst.  \textbf{39}(8),  1555--1571 (2020). \doi{10.1109/TCAD.2019.2915318}, \url{https://doi.org/10.1109/TCAD.2019.2915318}

\bibitem{CacheAttack}
Osvik, D.A., Shamir, A., Tromer, E.: {{Cache attacks and countermeasures: the case of AES}}. In: Topics in Cryptology--CT-RSA 2006: The Cryptographers’ Track at the RSA Conference 2006, San Jose, CA, USA, February 13-17, 2005. Proceedings. pp. 1--20. Springer (2006)

\bibitem{PesslP21}
Pessl, P., Prokop, L.: {Fault Attacks on CCA-secure Lattice KEMs}. {IACR} Trans. Cryptogr. Hardw. Embed. Syst.  \textbf{2021}(2),  37--60 (2021). \doi{10.46586/tches.v2021.i2.37-60}, \url{https://doi.org/10.46586/tches.v2021.i2.37-60}

\bibitem{Proos_Zalka_2003}
Proos, J., Zalka, C.: {Shor's discrete logarithm quantum algorithm for elliptic curves}. Quantum Inf. Comput.  \textbf{3}(4),  317--344 (2003). \doi{10.26421/QIC3.4-3}, \url{https://doi.org/10.26421/QIC3.4-3}

\bibitem{ParallelPC1}
Rajendran, G., Ravi, P., D'Anvers, J., Bhasin, S., Chattopadhyay, A.: {Pushing the Limits of Generic Side-Channel Attacks on LWE-based KEMs - Parallel {PC} Oracle Attacks on Kyber {KEM} and Beyond}. {IACR} Trans. Cryptogr. Hardw. Embed. Syst.  \textbf{2023}(2),  418--446 (2023). \doi{10.46586/tches.v2023.i2.418-446}, \url{https://doi.org/10.46586/tches.v2023.i2.418-446}

\bibitem{DBLP:journals/iacr/RaviCB22}
Ravi, P., Chattopadhyay, A., Baksi, A.: {{Side-channel and Fault-injection attacks over Lattice-based Post-quantum Schemes (Kyber, Dilithium): Survey and New Results}}. {IACR} Cryptol. ePrint Arch. p.~737 (2022), \url{https://eprint.iacr.org/2022/737}

\bibitem{Sidechannelattacks1}
Ravi, P., Roy, S.S., Chattopadhyay, A., Bhasin, S.: {Generic Side-channel attacks on CCA-secure lattice-based {PKE} and KEMs}. {IACR} Trans. Cryptogr. Hardw. Embed. Syst.  \textbf{2020}(3),  307--335 (2020), \url{https://doi.org/10.13154/tches.v2020.i3.307-335}

\bibitem{RSASig}
Razavi, K., Gras, B., Bosman, E., Preneel, B., Giuffrida, C., Bos, H.: {Flip Feng Shui: Hammering a Needle in the Software Stack}. In: Proceedings of the 25th USENIX Conference on Security Symposium. p. 1–18. SEC'16, USENIX Association, USA (2016)

\bibitem{OdedLecture}
Regev, O.: {Lecture notes: Lattices in computer science}, \url{https://cims.nyu.edu/~regev/teaching/lattices\_fall\_2009}

\bibitem{RSA}
Rivest, R.L., Shamir, A., Adleman, L.M.: {A Method for Obtaining Digital Signatures and Public-Key Cryptosystems}. Commun. {ACM}  \textbf{21}(2),  120--126 (1978). \doi{10.1145/359340.359342}, \url{http://doi.acm.org/10.1145/359340.359342}

\bibitem{seaborn2015exploiting}
Seaborn, M., Dullien, T.: {{Exploiting the DRAM rowhammer bug to gain kernel privileges}}. Black Hat  \textbf{15}, ~71 (2015)

\bibitem{CacheAttack2}
Settana, M., Naila, A., Yaseen, H., Huwaida, T.: {{Cache-Timing Attack against AES Crypto-Systems Countermeasure Using Weighted Average Masking Time Algorithm}}. Journal of Information Warfare  \textbf{15}(1),  104--114 (2016), \url{https://www.jstor.org/stable/26487484}

\bibitem{Shor_1994}
Shor, P.W.: {Algorithms for Quantum Computation: Discrete Logarithms and Factoring}. In: 35th Annual Symposium on Foundations of Computer Science, Santa Fe, New Mexico, USA, 20-22 November 1994. pp. 124--134. {IEEE} Computer Society (1994). \doi{10.1109/SFCS.1994.365700}, \url{https://doi.org/10.1109/SFCS.1994.365700}

\bibitem{ParallelPC}
Tanaka, Y., Ueno, R., Xagawa, K., Ito, A., Takahashi, J., Homma, N.: {Multiple-Valued Plaintext-Checking Side-Channel Attacks on Post-Quantum KEMs} (2022), \url{https://eprint.iacr.org/2022/940}

\bibitem{hammertime}
Tatar, A., Giuffrida, C., Bos, H., Razavi, K.: {Defeating Software Mitigations Against Rowhammer: {A} Surgical Precision Hammer}. In: Bailey, M., Holz, T., Stamatogiannakis, M., Ioannidis, S. (eds.) Research in Attacks, Intrusions, and Defenses - 21st International Symposium, {RAID} 2018, Heraklion, Crete, Greece, September 10-12, 2018, Proceedings. Lecture Notes in Computer Science, vol. 11050, pp. 47--66. Springer (2018). \doi{10.1007/978-3-030-00470-5\_3}, \url{https://doi.org/10.1007/978-3-030-00470-5\_3}

\bibitem{proof_shuffle}
Veyrat{-}Charvillon, N., Medwed, M., Kerckhof, S., Standaert, F.: Shuffling against side-channel attacks: {A} comprehensive study with cautionary note. In: Wang, X., Sako, K. (eds.) Advances in Cryptology - {ASIACRYPT} 2012 - 18th International Conference on the Theory and Application of Cryptology and Information Security, Beijing, China, December 2-6, 2012. Proceedings. Lecture Notes in Computer Science, vol.~7658, pp. 740--757. Springer (2012). \doi{10.1007/978-3-642-34961-4\_44}

\bibitem{XiaoZZT16}
Xiao, Y., Zhang, X., Zhang, Y., Teodorescu, R.: {One Bit Flips, One Cloud Flops: Cross-VM Row Hammer Attacks and Privilege Escalation}. In: Holz, T., Savage, S. (eds.) 25th {USENIX} Security Symposium, {USENIX} Security 16, Austin, TX, USA, August 10-12, 2016. pp. 19--35. {USENIX} Association (2016), \url{https://www.usenix.org/conference/usenixsecurity16/technical-sessions/presentation/xiao}

\bibitem{YaromF14}
Yarom, Y., Falkner, K.: {{FLUSH+RELOAD:} {A} High Resolution, Low Noise, {L3} Cache Side-Channel Attack}. In: Fu, K., Jung, J. (eds.) Proceedings of the 23rd {USENIX} Security Symposium, San Diego, CA, USA, August 20-22, 2014. pp. 719--732. {USENIX} Association (2014), \url{https://www.usenix.org/conference/usenixsecurity14/technical-sessions/presentation/yarom}

\end{thebibliography}
\appendix
\section*{Supplementary material}

\section{Rowhammer}\label{app:rowhammer}

Rowhammer is a phenomenon observed in dynamic random-access memory (DRAM) where repeated access to a specific row can result in bit flips occurring in neighboring rows \cite{Flippingbit1}. This happens because the capacitors in neighboring rows, responsible for storing bit values, discharge slightly due to parasitic currents when the row's word line is activated if this discharge happens frequently enough to lower the voltage below the "charged" threshold before the regular DRAM refresh, which typically occurs every 64ms, the logical value of the bit can flip. There are two types of methods for row hammering: single-sided rowhammering and double-sided rowhammering.

\noindent\textbf {Single-sided Rowhammering:}
Single-sided rowhammering is a technique where an attacker repeatedly accesses memory rows on only one side of a target row without accessing the rows above or below within the same bank. By rapidly accessing specific rows, the attacker aims to induce electrical interference and disturb neighboring cells, potentially causing bit flips and altering data stored in adjacent rows. This technique relies on the inherent electrical interactions between closely located memory cells, taking advantage of the sensitivity of DRAM cells to repeated accesses.

\noindent\textbf { Double-sided Rowhammering: }
Double-sided rowhammering is a more aggressive variant of the rowhammering technique. In this approach, an attacker repeatedly accesses memory rows above and below a target row within the same bank. By accessing these rows simultaneously, the attacker intensifies the electrical interactions and disturbances within the DRAM cells. This increases the likelihood of inducing bit flips in the target and adjacent rows. Double-sided row hammer leverages the interplay of electrical charges in closely positioned memory cells to exploit the vulnerability of DRAM and manipulate data stored in memory.
\section{Fujisaki-Okamoto (FO) transformation} \label{app:fo}
We use this FO transformation proposed by \cite{Jiang2017} to construct a CCA secure key-encapsulation mechanism (KEM) over the CPA secure \texttt{LPR.PKE} scheme. The algorithms of this KEM are shown in Figure~\ref{fig:fo-kem}.  
The KEM algorithm contains three algorithms: (i) key generation (\texttt{KEM.KeyGen}), (ii) encapsulation (\texttt{KEM.Encaps}), and (iii) decapsulation (\texttt{KEM.Decaps}). The algorithm \texttt{KEM.KeyGen} produces a public key $pk$ and a secret key $sk'$ by employing the \texttt{LPR.PKE.KeyGen} algorithm. In this context, $sk$ refers to the secret key generated through the \texttt{LPR.PKE.KeyGen} algorithm, $z$ is a random bit string of length $n$, and $\mathcal{H}$ is a hash function. The secret key $sk'$ is computed by concatenating $sk$, $pk$, $\mathcal{H}(pk)$, and $z$. The \texttt{KEM.Encaps} algorithm takes the public key $pk$ as input and generates a random message bit string $m$ of length $n$. Then it uses the hash function $\mathcal{G}$ to compute $K'$ and a random coin string $r$. After that, it encrypts the message $m$ using the \texttt{LPR.PKE.Enc} algorithm with public key $pk$, message $m$, and random coin string $r$ to produce the ciphertext $ct$. Finally, it applies a function $\mathcal{F}$ to $K'$ and the ciphertext $ct$ to produce the shared key $K$. The \texttt{KEM.Decaps} algorithm takes the ciphertext $ct$ and secret key $sk'$ as inputs. It first decrypts the ciphertext using the \texttt{LPR.PKE.Dec} algorithm with secret key $sk$ and ciphertext $ct$ to produce the decrypted message $m'$. It then re-encrypts the message $m'$ using the \texttt{LPR.PKE.Enc} algorithm with public key $pk$ and random coin string $r$ to produce the ciphertext $c$. It then checks whether $ct$ and $c$ are equal. If they are, it applies the function $\mathcal{F}$ to $K'$ and the hash of the ciphertext $\mathcal{H}(ct)$ to produce the shared key $K$. Otherwise, it applies the function $\mathcal{F}$ to the random bit string $z$ and the hash of the ciphertext $\mathcal{H}(ct)$ to produce an invalid shared key $K$. Finally, it returns the shared key $K$.
\end{document}